\documentclass[twocolumn,journal]{IEEEtran}

\IEEEoverridecommandlockouts

\usepackage{bm}
\usepackage{color}
\usepackage{graphicx}
\usepackage{amsmath}
\usepackage{amssymb}
\usepackage{algorithm}
\usepackage{algorithmic}
\usepackage{ulem}
\usepackage{lineno}
\usepackage{lettrine}
\usepackage{subfigure}
\usepackage{epstopdf}
\usepackage{stfloats}
\usepackage{multirow}
\usepackage{booktabs}
\usepackage{array}
\usepackage{amsthm}
\usepackage{caption}
\usepackage{subfigure}
\usepackage{setspace}
\usepackage{makecell}
\usepackage{yhmath}
\allowdisplaybreaks[4]

\newcommand{\be}{\begin{equation}}
	\newcommand{\ee}{\end{equation}}
\newcommand{\bea}{\begin{eqnarray}}
	\newcommand{\eea}{\end{eqnarray}}
\newcommand{\ba}{\begin{array}}
	\newcommand{\ea}{\end{array}}

\newcounter{mytempeqncnt}
\def\BibTeX{{\rm B\kern-.05em{\sc i\kern-.025em b}\kern-.08em
    T\kern-.1667em\lower.7ex\hbox{E}\kern-.125emX}}
\usepackage{balance}

\begin{document}
\title{Joint BS-RIS-User Association and Beamforming Design for RIS-assisted Cellular Networks}

\author{Sifan Liu,
        Rang Liu,~\IEEEmembership{Graduate Student Member,~IEEE,}
        Ming Li,~\IEEEmembership{Senior Member,~IEEE,}
        Yang Liu,~\IEEEmembership{Member,~IEEE,}
        and Qian Liu,~\IEEEmembership{Member,~IEEE}
\thanks{ S. Liu, R. Liu, M. Li, and Y. Liu are with the School of Information and Communication
Engineering, Dalian University of Technology, Dalian 116024, China (e-mail:
sifanliu@mail.dlut.edu.cn; liurang@mail.dlut.edu.cn; mli@dlut.edu.cn; yangliu$\_$613@dlut.edu.cn).}
\thanks{ Q. Liu is with the School of Computer Science and Technology, Dalian University
of Technology, Dalian 116024, China (e-mail: qianliu@dlut.edu.cn).}}

%\markboth{Journal of \LaTeX\ Class Files,~Vol.~18, No.~9, September~2020}%
%{How to Use the IEEEtran \LaTeX \ Templates}

\maketitle
\thispagestyle{empty}
\begin{abstract}
Reconfigurable intelligent surface (RIS) is a revolutionary technology for sixth-generation (6G) networks owing to its ability to manipulate wireless environments.
As a frequency-selective device, RIS can only effectively shape the propagation of signals within a certain frequency band.
Due to this frequency-selective property, the deployment of RIS in cellular networks will introduce a complicated base station (BS)-RIS-user association issue since adjacent BSs operate at different frequency bands.
In this paper, with the consideration of the frequency-selective characteristics of RIS, we aim to jointly optimize BS-RIS-user association, active beamforming at BSs, and passive beamforming of RIS to maximize the sum-rate of a RIS-assisted cellular network.
We first leverage $l_0$-norm to efficiently integrate BS-RIS-user association with active and passive beamforming.
Then, we adopt fractional programming (FP) and block coordinate descent (BCD) methods to deal with logarithmic and fractional parts and decouple the joint association and beamforming design problem into several sub-problems.
Efficient algorithms which combine $l_0$-norm approximation, majorization-minimization (MM), and alternating direction method of multipliers (ADMM) are developed to alternately solve the sub-problems.
Extensive simulation results illustrate the importance of BS-RIS-user association optimization in RIS-assisted cellular networks and verify the effectiveness of the proposed joint association and beamforming design algorithm.
\end{abstract}


\section{Introduction}
Recently, with the rapid deployment of commercial fifth-generation (5G) wireless networks worldwide, researchers shift their interests to future sixth-generation (6G) systems, which require higher data rate and network capacity.
However, the existing technologies in 5G, e.g., massive multiple-input multiple-output (MIMO), millimeter wave (mmWave) communications, and ultra-dense heterogeneous network have reached their inherent limitations on transmission rate, spectrum/energy efficiency, etc. %\cite{W. Saad}.
%It is difficult for these technologies to meet the demanding even though with high energy consumptions and hardware costs \cite{W. Saad}, \cite{C. Pan}.
Thereby, innovative technologies are needed to achieve radical changes.

In conventional networks, the wireless propagation environment is regarded as a randomly varied uncontrollable factor, which cannot be optimized in the network design.
Recently, an innovative reconfigurable intelligent surface (RIS) technology is introduced to radically change this conventional perception.
RIS is deemed as a revolutionary and promising technology for 6G because of its ability to reconfigure the wireless propagation environment \cite{C. Pan}-\cite{E. Basar}. %\cite{Q. Wu1}
To be specific, RIS is a meta-surface consists of a large number of low-cost passive reflecting elements.
Each element can be independently controlled to collaboratively achieve passive reflection beamforming by adjusting the amplitude/phase-shift of incident signals.
Owing to the superior ability of shaping wireless propagation environment, the deployment of RIS can provide new degrees of freedom (DoFs) for the design of wireless networks, which can significantly improve the performance of networks.

Benefiting from the simple and passive hardware architecture, RIS features low energy consumption, low hardware cost, and easy installation, etc.
These appealing advantages stimulate many research efforts of designing RIS-assisted wireless communication systems using different metrics, including maximizing spectral efficiency \cite{X. Yu}, maximizing energy efficiency \cite{C. Huang}, \cite{G. Lee}, maximizing sum-rate of all users \cite{G. Zhou}-\cite{H. Guo}, maximizing received power \cite{P. Wang}, and minimizing transmit power \cite{Q. Wu3}-\cite{M.-M. Zhao}, etc.
In addition, due to passive and massive RIS reflecting elements, channel state information (CSI) acquisition is challenging, which has triggered the research on efficient channel estimation for RIS-assisted wireless systems \cite{B. Zheng2}-\cite{L. Wei}.
Moreover, RIS is also applied in combination with other emerging technologies, such as non-orthogonal multiple-access (NOMA) \cite{B. Zheng3}, \cite{T. Hou}, simultaneous wireless information and power transfer (SWIPT) \cite{Q. Wu5}, \cite{C. Pan2}, symbol-level precoding (SLP) \cite{R. Liu1}, \cite{R. Liu2}, holographic MIMO (HMIMO) \cite{C. Huang2}, integrated sensing and communications (ISAC) \cite{X. Song}-\cite{Z. Zhu}, etc.

Most existing works on RIS-assisted systems consider relatively simple scenarios with only one transmitter or base station (BS).
In order to accelerate the deployment of RIS in realistic cellular networks, the design of RIS in multi-cell systems is also a worth studying problem.
Moreover, two special and crucial issues are needed to be considered for the RIS-assisted cellular network.
Firstly, some existing works illustrated the frequency-selective property of RIS. By considering the hardware complexity and cost in practical implementations, this property can be approximately modeled as: Each element of RIS can only effectively provide tunable phase-shift for signals within a certain frequency band, while generates almost fixed phase-shift for signals at other frequency bands \cite{H. Li}-\cite{W. Cai2}. Notice that the performance loss caused by this simplified RIS model is verified to be negligible \cite{W. Cai2}.
%Considering this frequency-selective property of RIS, in the cellular network where the adjacent BSs usually operate at different frequency bands, RIS can only optimize its phase-shifts to effectively assist one BS operating at a certain frequency band, while its phase-shifts are almost fixed for signals from other BSs.
Considering that the adjacent BSs usually operate at different frequency bands in the cellular network, the practical RIS with this frequency-selective property can only optimize its phase-shifts to effectively assist one BS operating at a certain frequency band, while its phase-shifts are almost fixed for signals from other BSs.
%Firstly, as a frequency-selective device, each element of RIS can only effectively provide tunable phase-shifts for signals within a certain frequency band, while generates almost fixed phase-shifts for signals at other frequency bands \cite{H. Li}-\cite{W. Cai2}. % \cite{W. Cai},
%Thus, in the cellular network where the adjacent BSs usually operate at different frequency bands, RIS can only optimize its phase-shifts to effectively assist one BS operating at a certain frequency band, while its phase-shifts are almost fixed for signals transmitted from other BSs.
This fact means that the BS-RIS association problem deserves to be carefully studied in the RIS-assisted multi-cell system and effective association algorithms need to be developed for this new BS-RIS association problem.
Secondly, %in the conventional multi-cell systems, BS-user association problem is usually an important issue, which plays a pivotal role in enhancing the system  performance \cite{D. Liu}, \cite{T. Van Chien}. However, in the cellular system
the deployment of RIS introduces a different and more complicated BS-user association problem than conventional multi-cell systems, because the BS-user link relies not only on the direct channels, but also on the BS-RIS-user cascaded channels.
Moreover, since RIS can reflect the signals from all associated/un-associated BSs, the BS-user association in all cells will be affected.
This fact reveals that the traditional user association algorithms cannot be directly adopted in RIS-assisted systems, which inspires us to investigate new BS-user association algorithms.

Some researchers have noticed the impact of deploying RIS on the user association as well as the RIS association in cellular networks and started to investigate these BS-user \cite{E. M. Taghavi}-\cite{D. Zhao2}, BS-RIS \cite{D. Jin}, RIS-user \cite{Y. Huang}, \cite{W. Mei1}, or BS-RIS-user \cite{W. Mei2}, \cite{S. Liu} association problems in RIS-assisted systems.
In \cite{E. M. Taghavi}, the authors proposed a load balancing BS-user association scheme for multi-RIS assisted multi-cell system, in which each BS can be assisted by one RIS.
%This fact will impact the optimization results of the passive beamforming at RISs, thus affecting the system performance.
In \cite{P. Han}, a BS-user association and passive beamforming optimization problem based on statistical CSI was investigated for a multi-BS multi-RIS heterogeneous network.
The authors in \cite{D. Zhao2} presented a joint power allocation and BS-user association design problem for mmWave systems.
However, the authors of these works \cite{E. M. Taghavi}-\cite{D. Zhao2} assumed that the BS-RIS association is fixed, which is not realistic.
Moreover, they ignored that the RIS can reflect the signals of all BSs rather than only the associated BS. %several RISs are deployed to assist only one BS in the multi-BS network and , which also disregarded the feature that RIS will reflect the signals from other BSs.
The authors in \cite{D. Jin} focused on the BS-RIS association and beamforming design problem and proposed a multi-agent deep reinforcement learning-based association scheme. %in a multi-BS multi-RIS system
The authors in \cite{Y. Huang} considered a RIS-user association and cascaded channel estimation problem, while the authors in \cite{W. Mei1} presented the RIS-user association problem with fixed BS-user association.
The authors in \cite{W. Mei2} aimed to maximize the utility of users by optimizing the BS-RIS-user association and passive beamforming in a multi-BS multi-RIS narrow-band system.
Nevertheless, the authors in \cite{W. Mei2} did not take the active beamforming optimization into consideration. Furthermore, they only designed the passive beamforming by aligning the angles of direct and cascade channels when the RIS is associated with one BS, which ignores the impact of the reflecting signals from other BSs.
In \cite{S. Liu}, a BS-RIS-user association and beamforming design problem was investigated for a cellular network, in which the power allocation of users in each cell is not considered.

These existing works reviewed above only investigate part of joint BS-RIS-user association, active beamforming, and passive beamforming design, but ignore that the BS-RIS-user association and active/passive beamforming optimization problems are tightly interrelated and inseparable. %in practice. %, which is more realistic but intractable because of the highly coupled variables.
More importantly, the frequency-selective property of RIS has not been considered in most of the existing literatures.
These facts mean that most of the existing association algorithms cannot be directly utilized in the joint BS-RIS-user association and beamforming design for practical RIS-assisted systems. Thus, the association and beamforming design in practical RIS-assisted systems is worth to be investigated, which is more practical, comprehensive, and realistic compared with the existing works.
Motivated by above discussion, in this paper we consider a practical RIS-assisted multi-cell wireless network and focus on the BS-RIS-user association as well as the active and passive beamforming optimization.
Our main contributions are summarized as follows:
\begin{itemize}
\item Different from most of existing works which only consider part of the BS-RIS-user association problem with an ideal RIS model, a more comprehensive and realistic practical RIS-assisted multi-cell network is modeled, in which we concern about the joint design of BS-RIS-user association, the active beamforming of BSs, and the passive beamforming of RIS.
It is worth mentioning that we adopt the practical RIS reflecting model which describes the frequency-selective characteristics of RIS \cite{H. Li}-\cite{W. Cai2}.
    The sum-rate maximization problem is then formulated, which aims to maximize the sum-rate of all users in the cellular network subject to the power budgets of BSs, the association constraints, and the restrict of RIS elements.
\item To facilitate the joint association and beamforming design, we utilize $l_0$-norm to combine the active beamforming with BS-user association and integrate the passive beamforming with BS-RIS association.
    %The is used to represent the association in the optimization problem., which will make the problem more tractable
    Then, in order to solve the non-convex non-smooth joint association and beamforming optimization problem, we first adopt the fractional programming (FP) and block coordinate descent (BCD) method to deal with logarithmic and fractional parts, and then decouple the problem into several sub-problems.
    Efficient algorithms which combine $l_0$-norm approximation, majorization-minimization (MM), and alternating direction method of multipliers (ADMM) are developed to alternately solve the sub-problems.
\item Finally, extensive simulation results are provided to verify the importance of the joint association and beamforming design as well as the effectiveness of our proposed algorithm.
    To be specific, the proposed joint association and beamforming design algorithm facilitates load balancing between BSs, which is beneficial to the utilization of resources. Moreover, %the performance gap between the proposed algorithm and the comparison algorithm also verified that the joint BS-RIS-user association in RIS-assisted cellular network will promote the performance of system.
    %To be specific,the deployment of RIS in cellular network can expand cell edge. Moreover,
    our proposed algorithm has significant performance improvement compared with benchmark algorithms, which demonstrates the advancement of joint BS-RIS-user association design.
\end{itemize}

%The rest of this paper is organized as follows.
%Section II mainly presents the system model and the problem formulation of the association and beamforming optimization problem.
%The proposed FP-MM-ADMM based association and beamforming optimization algorithm is introduced in detail in section III.
%Simulation results are shown in section IV. Finally, the conclusions are provided in section V.

\textit{Notations}: Boldface lower-case, boldface upper-case, and Italic letters indicate column vectors, matrices, and scalars, respectively.
$\mathbb{C}$ and $\mathbb{R}$ denote the sets of complex and real numbers, respectively.
$\jmath$ presents the imaginary unit.
For a complex valued vector $\boldsymbol{a}$, $\|\boldsymbol{a}\|_0$ and $\|\boldsymbol{a}\|_2$ denote its $l_0$-norm and $l_2$-norm, respectively.
$\mathbf{A}^{T}$ and $\mathbf{A}^{H}$ denotes the transpose and conjugate transpose for matrix $\mathbf{A}$, respectively.
$\mathbb{E}\{\cdot\}$ and $\mathfrak{Re}\{\cdot\}$ indicate statistical expectation and the real part of a complex number, respectively.
$\mathbf{I}_M$ presents an $M \times M$ identity matrix. %$\odot$ denotes the Hadamard product.
$\oslash$ denotes the element-wise division.
\section{System Model and Problem Formulation}

\subsection{System Model}
\begin{figure}[!t]
	\centering
	\includegraphics[width = 3.5 in]{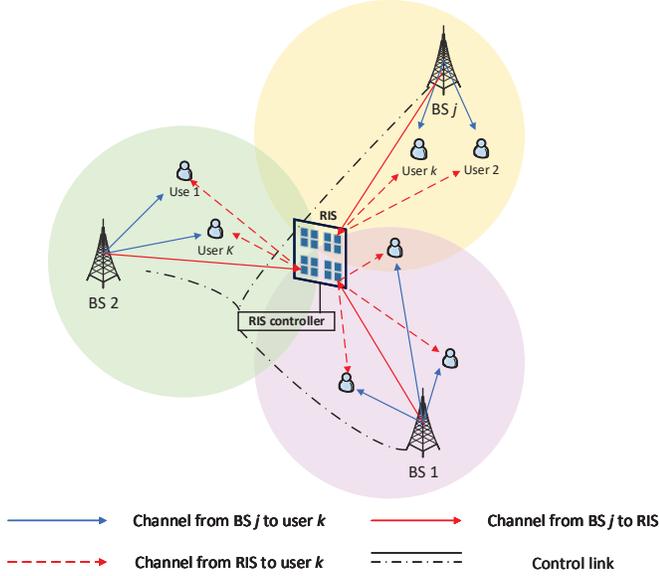}
	\caption{RIS-assisted multi-cell MU-MISO wireless communication network.}
	\label{fig:system}
\end{figure}
We consider a RIS-assisted multi-cell wireless network as shown in Fig. \ref{fig:system}, where $K$ single-antenna users are served by $J$ BSs operating at different frequency bands.
Each BS is equipped with $M$ antennas and can simultaneously transmit independent data streams to a maximum of $K$ users.
Furthermore, a RIS, which consists of $N$ reflecting elements, is deployed to assist the communications between BSs and their associated users.
Thus, each cell can be deemed as a RIS-assisted multi-user multiple-input single-output (MU-MISO) system.
For brevity, let $\mathcal{J} = \{1,2,...,J\}$, $\mathcal{K} = \{1,2,...,K\}$, and $\mathcal{N} = \{1,2,...,N\}$ denote the sets of BSs, users, and RIS reflecting elements, respectively.
%In addition, we assume that the RIS is controlled by BSs via wireless or wired links and the CSI of all channels are perfectly known by BSs with the aid of existing efficient channel estimation algorithms for RIS-assisted communications \cite{B. Zheng2}-\cite{L. Wei}.
In addition, we assume that the RIS is controlled by BSs via dedicated wireless or wired links. Furthermore, for the channel estimation in RIS-assisted systems, many effective and efficient frameworks/algorithms have been proposed, which can achieve accurate channel estimation with low pilot overhead \cite{B. Zheng2}-\cite{L. Wei}. Thus, we also assume that the CSI of all channels are perfectly estimated and known by BSs.

Recent works illustrate that the reflecting elements of a practical RIS can be approximately considered only providing tunable phase-shifts for the signals within a certain frequency band (the frequency band can be modified by adjusting the capacitance of RIS elements), and having almost fixed ($0$ or $2\pi$) phase-shifts for the signals at other frequency bands \cite{H. Li}-\cite{W. Cai2}. %\cite{W. Cai},
Meanwhile, this simplified practical model can significantly reduce the computational complexity for the RIS passive beamforming design compared with the practical RIS which is modeled by microwave theory, while the performance loss caused by the approximation is verified to be negligible \cite{W. Cai2}.
This important frequency-selective feature indicates that the deployed RIS can effectively change the propagation of signals for only one BS, since adjacent BSs operate at different frequency bands.
In other words, the RIS can only effectively serve one BS by adaptively manipulating the wireless propagation between this BS and its users.
Thus, BS-RIS association with the consideration of the frequency-selective property of RIS is a new but crucial problem in the RIS-assisted network.
To efficiently express the BS-RIS association as well as the passive beamforming of RIS, we first assume that the phase-shifts of RIS are fixed as $0$ for the signals at other un-associated BSs' operating frequency bands, since phase-shifts $0$ or $2\pi$ will generate the same reflecting coefficients.
By defining the phase-shifts of RIS for BS-$j$ as $\boldsymbol{\theta}_{j}\triangleq[\theta_{j,1},\theta_{j,2},...,\theta_{j,N}]^T$, the BS-RIS association and passive beamforming can be combined.
%because of the frequency-selective characteristic of RIS
%If $r_j=1$, RIS will assist BS-$j$ by adaptively tune the phase-shifts for the signals at the frequency band of BS-$j$; otherwise, $r_j=0$, BS-$j$ is not assisted by RIS, i.e., the phase-shifts for the signal of BS-$j$ are not tunable and fixed as $0/2\pi $.
%By utilizing the frequency-selective characteristic of RIS in BS-RIS association, we can integrate the BS-RIS association and phase-shifts into vector $\boldsymbol{\theta}_{j}$.
Specifically, $\boldsymbol{\theta}_{j} = \mathbf{0}_N$ indicates that BS-$j$ is not assisted by RIS, i.e., the phase-shifts are not tunable and fixed as zeros; otherwise, $\boldsymbol{\theta}_{j} \ne \mathbf{0}_N$ represents that RIS will assist BS-$j$ by adaptively tuning the phase-shifts for the signals at the frequency band of BS-$j$.
%Thus, the BS-RIS association and passive beamforming are combined to the vector $\boldsymbol{\theta}_{j} $.
Then, the passive beamforming matrix of RIS for BS-$j$ can be represented by $\boldsymbol{\Phi}_{j}\triangleq\operatorname{diag}(\boldsymbol{\varphi}_{j} )$, where $\boldsymbol{\varphi}_{j}\triangleq e^{\jmath\boldsymbol{\theta}_{j}}=[e^{\jmath\theta_{j,1}},e^{\jmath\theta_{j,2}},...,e^{\jmath\theta_{j,N}}]^T$.

According to the frequency-selective characteristics of RIS, it can only assist at most one BS and serve its associated users in the cellular system.
This feature in BS-RIS association can be easily expressed as the constraint $\big\|[\left\|\boldsymbol{\theta}_{1}\right\|_2^2,...,\left\|\boldsymbol{\theta}_{J}\right\|_2^2]\big\|_0 = \sum_{j \in \mathcal{J}} \big\|\left\|\boldsymbol{\theta}_{j}\right\|_2^2\big\|_0 = 1$ by utilizing $l_0$-norm, which represents the number of non-zero elements in vector $[\left\|\boldsymbol{\theta}_{1}\right\|_2^2,...,\left\|\boldsymbol{\theta}_{J}\right\|_2^2]^T$.
In order to convert the problem into a more common form, we prefer to use the reflecting coefficient $\boldsymbol{\varphi}_{j}$ instead of the angle $\boldsymbol{\theta}_{j}$ in this constraint.
%It is obvious that constraint \eqref{eq:constraint_theta1} can be replaced by $|\varphi_{j,n}|=1, ~ \forall j \in \mathcal{J},~\forall n \in \mathcal{N}$.
Thus, based on the definition $\boldsymbol{\varphi}_{j}\triangleq e^{\jmath\boldsymbol{\theta}_{j}}$, the BS-RIS association constraint can be reformulated as
\begin{equation}
\label{eq:ln}
\begin{aligned}
\sum_{j \in \mathcal{J}} \big\|\|\boldsymbol{\theta}_{j}\|_2^2\big\|_0
\overset{\text{(a)}}=&\sum_{j \in \mathcal{J}}\left\|\left\|-\jmath\ln(\boldsymbol{\varphi}_{j})\right\|_2^2\right\|_0\\
\overset{\text{(b)}}=& \sum_{j \in \mathcal{J}}\left\|\left\|\ln(\boldsymbol{\varphi}_{j})\right\|_2^2\right\|_0 = 1,
\end{aligned}
\end{equation}
where $\ln(\boldsymbol{\varphi}_{j})$ represents an element-wise natural logarithmic operation.
The equality (a) holds according to the definition of complex logarithmic function, i.e., $a = \log (b) = \log |b| + \jmath (\arg(b) + 2k\pi),~k = 0,\pm1,\pm2,...$, in which $a$ and $b$ are complex numbers. Then, with the unit modulus of $\varphi_{j,n}$, we can obtain the phase-shift of $\varphi_{j,n}$ by logarithmic operations. Notice that we only concern about the principal value $\boldsymbol{\theta}_{j} = -\jmath\ln(\boldsymbol{\varphi}_{j})$, which lies in the definition domain $\left[-\pi,\pi\right)$. Equality (b) holds since $\jmath^2 = -1$.

On the other hand, the association between BSs and users is also significantly affected by the deployment of RIS due to additional reconfigurable BS-RIS-user cascaded channel.
Considering that the conventional BS-user association algorithms cannot be directly applied for the RIS-assisted network, we should carefully elaborate on the BS-user association problem for the considered system. % and the BS-user association problem should be re-investigated.
%To better represent the BS-user association in the system model and problem formulation, similar to the BS-user association variable $r_{j}$, we also define user association variable $z_{j,k} \in \{0,1\}, ~ \forall j \in \mathcal{J},~ \forall k \in \mathcal{K} $. Specifically, $z_{j,k} = 1$ indicates that user-$k$ is served by BS-$j$; otherwise, $z_{j,k} = 0$ implies that BS-$j$ will not transmit any information to user-$k$. Moreover, we assume that each user is served by only one BS, which can be expressed as the constraint $\sum_{j \in \mathcal{J}} z_{j,k} = 1,~ \forall k \in \mathcal{K}$.
Similar to the integration of BS-RIS association and passive beamforming, we combine the BS-user association with the active beamforming by utilizing $l_0$-norm.
%To better represent the BS-user association,
Specifically, we first define $\mathbf{w}_{j,k} \in \mathbb{C}^{M \times 1}$ as the beamforming vector at BS-$j$ for user-$k$. %, which is meaningless and equals to zero if user-$k$ is not served by BS-$j$.
When user-$k$ is not served by BS-$j$, active beamforming $\mathbf{w}_{j,k}=\mathbf{0}_M$; otherwise, if user-$k$ is associated with BS-$j$, $\mathbf{w}_{j,k}\ne \mathbf{0}_M$ and can be optimized.
%Thus, the BS-user association and active beamforming can be integrated into the vector $\mathbf{w}_{j,k}$.
Since we assume that each user is served by only one BS, the BS-user association can be expressed as the constraint $\big\|[\left\|\mathbf{w}_{1, k}\right\|_2^2,...,\left\|\mathbf{w}_{J, k}\right\|_2^2]\big\|_0 = \sum_{j \in \mathcal{J}}\big\|\left\|\mathbf{w}_{j, k}\right\|_2^2\big\|_0 = 1,~ \forall k$.

Denote $s_{j,k}$ as the symbol transmitted from BS-$j$ to user-$k$, which satisfies $\mathbb{E}\{s_{j,k}s_{j,k}^*\}=1$ and $\mathbb{E}\{s_{j,k}s_{j,i}^*\}=0,~\forall i \ne k$.
%Moreover, we denote $\mathbf{w}_{j,i} \in \mathbb{C}^{M \times 1}$ as the beamforming vector from BS-$j$ to user-$i$.
We should emphasize that the symbol $s_{j,k}$ is meaningful only when BS-$j$ is associated to user-$k$. %, i.e., $z_{j,i} = 1$.and $\mathbf{w}_{j,i}$ are
Thus, by denoting $\mathbf{h}_{\text{d},j,k}\in \mathbb{C}^{M \times 1}$, $\mathbf{G}_{j} \in \mathbb{C}^{N \times M}$, and $\mathbf{h}_{\text{r},k} \in \mathbb{C}^{N \times 1}$, $\forall j,~\forall k $ as the channels from BS-$j$ to user-$k$, from BS-$j$ to RIS, and from RIS to user-$k$, respectively, the received signal of user-$k$ from the BS-$j$ can be written as
\begin{equation}
\label{eq:received-signal}
y_{j,k}=\left\{\begin{array}{ll}
 \sum \limits_{i \in \mathcal{K} } \widetilde{\mathbf{h}}_{j,k}^{H}\mathbf{w}_{j,i} s_{j,i}+n_{k}, & \text{BS-}j \text{ serves user-}k, \\
0, & \text{otherwise},
\end{array}\right.
\end{equation}
in which we define $\widetilde{\mathbf{h}}_{j,k}^{H}\triangleq\mathbf{h}_{\text{d}, j, k}^{H}+\mathbf{h}_{\text{r}, k}^{H} \boldsymbol{\Phi}_{j} \mathbf{G}_{j}$ as the equivalent compound channel from BS-$j$ to user-$k$ for brevity.
$n_{k} \sim \mathcal{CN} (0, \sigma^{2}_k)$ is complex additive white Gaussian noise with variance $\sigma^{2}_k$ and zero mean.

Then, the signal-to-interference-plus-noise ratio (SINR) of user-$k$ can be represented as
\begin{equation}
\label{eq:SINR}
\begin{aligned}
\mathrm{SINR}_{j,k}=& \frac{|\widetilde{\mathbf{h}}_{j,k}^{H} \mathbf{w}_{j, k}|^{2}}{\sum\limits_{i \in \mathcal{K}, i \neq k}|\widetilde{\mathbf{h}}_{j,k}^{H} \mathbf{w}_{j, i}|^{2}+\sigma^{2}_k},~\forall j ,~\forall k .
\end{aligned}
\end{equation}
Hence, the achievable rate of user-$k$ can be calculated by
\begin{equation}
\label{eq:rate}
	R_{j,k} = \log_{2} \left(1+\mathrm{SINR}_{j,k}\right), ~\forall j ,~\forall k .
\end{equation}
Similar to the transmitted symbol $s_{j,k}$, $\mathrm{SINR}_{j,k}$ and $R_{j,k}$ have meaningful values only when user-$k$ is served by BS-$j$; otherwise, they are meaningless and their values are all zeros.
\subsection{Problem Formulation}
We aim to maximize the achievable sum-rate by jointly optimizing the active beamforming $\mathbf{w}_{j,k}$ and passive beamforming $\boldsymbol{\varphi}_{j}$, which also indicate BS-user association and BS-RIS association, respectively.
Thus, the sum-rate maximization problem is formulated as
\begin{subequations}
\label{eq:problem_original_zeronorm}
\begin{align}
\mathop{\max}\limits_{\substack{\mathbf{w}_{j,k},\boldsymbol{\varphi}_{j}\\ \forall j \in \mathcal{J}, \forall k \in \mathcal{K}}} \;
& \sum_{j \in \mathcal{J}}\sum_{k \in \mathcal{K}} R_{j,k} \label{eq:original function}\\
\text {s.t.} \quad
& \sum_{k \in \mathcal{K}} \left\|\mathbf{w}_{j,k}\right\|_2^{2} \le P_{j}, ~ \forall j, \label{eq:power constraint}\\
& \sum_{j \in \mathcal{J}}\left\|\left\|\mathbf{w}_{j, k}\right\|_2^2\right\|_0  = 1, ~ \forall k , \label{eq:BS constraint}\\
& |\varphi_{j,n}|=1, ~ \forall j ,~ \forall n, \label{eq:angle constraint}\\	
& \sum_{j \in \mathcal{J}}\left\|\left\|\ln(\boldsymbol{\varphi}_{j})\right\|_2^2\right\|_0 = 1, \label{eq:theta constraint}
\end{align}
\end{subequations}where $P_{j}$ is the maximum transmit power of BS-$j$.
Constraint \eqref{eq:BS constraint} indicates that each user can only be served by one BS, and constraint \eqref{eq:theta constraint} implies that RIS can serve only one BS.  %during a coherent time interval.

The non-smooth non-convex problem \eqref{eq:problem_original_zeronorm} is difficult to solve because of the non-convex objective function \eqref{eq:original function}, non-smooth non-convex constraints \eqref{eq:BS constraint}, \eqref{eq:theta constraint}, unit-modulus constraint \eqref{eq:angle constraint}, and highly-coupled variables.
In order to tackle these difficulties, we propose an FP-MM-ADMM based joint association and beamforming design algorithm.

\section{Association and Beamforming Design}
In this section, we present the proposed FP-MM-ADMM based joint association and beamforming design algorithm.
To be specific, we utilize the FP method to convert the logarithmic and fractional objective function into a more tractable form.
Then, we adopt the BCD method to decompose the coupled variables and alternately solve the three resulting sub-problems, i.e., 1) the update of auxiliary variables, 2) the joint BS-user association and active beamforming optimization problem, and 3) the joint BS-RIS association and passive beamforming optimization problem.

\subsection{Objective Function Transformation}
In this subsection, we attempt to deal with the non-convex objective function \eqref{eq:original function} by FP method proposed in \cite{K. Shen}.
We first use the Lagrangian dual transform to handle the logarithmic form. By introducing the dual variable $\boldsymbol{\tau} \triangleq [\tau_{1,1},...,\tau_{j,k},...,\tau_{J,K}]^T$, the objective function \eqref{eq:original function} is equivalently written as
\begin{equation}
\label{eq:objective function f3}
\begin{aligned}
\sum_{j \in \mathcal{J}}& \sum_{k \in \mathcal{K}}\log _2\left(1+\tau_{j, k}\right)
- \frac{1}{\ln 2}\sum_{j \in \mathcal{J}} \sum_{k \in \mathcal{K}} \tau_{j, k}\\
&+\frac{1}{\ln 2}\sum_{j \in \mathcal{J}} \sum_{k \in \mathcal{K}} \frac{\left(1+\tau_{j,k}\right) |\widetilde{\mathbf{h}}_{j,k}^{H} \mathbf{w}_{j, k}|^{2}}{\sum\limits_{i \in \mathcal{K}}|\widetilde{\mathbf{h}}_{j,k}^{H} \mathbf{w}_{j, i}|^{2}+\sigma^{2}_k}.
\end{aligned}
\end{equation}%f_{3}(\mathbf{w}_{j,k},\boldsymbol{\theta}_{j},\boldsymbol{\tau})=&
Then, to deal with the fractional part, we adopt the quadratic transform by further introducing another auxiliary variable $\mathbf{q}\triangleq [q_{1,1},...,q_{j,k},...,q_{J,K}]^T$, which is used to decouple the numerators and denominators.
By denoting $\mathbf{w}\triangleq [\mathbf{w}_{1,1}^H,...,\mathbf{w}_{j,k}^H,...,\mathbf{w}_{J,K}^H]^H$ and $\boldsymbol{\varphi}\triangleq [\boldsymbol{\varphi}_{1}^H,...,\boldsymbol{\varphi}_{J}^H]^H$% as $\mathbf{w}_{j,k},~\forall j \in \mathcal{J}, ~\forall k \in \mathcal{K}$ and $\boldsymbol{\varphi}_{j},~\forall j \in \mathcal{J}$ for brevity, respectively
, the objective function \eqref{eq:objective function f3} can be further converted into \eqref{eq:objective function f4} shown at the top of next page.
\begin{figure*}[!t]
\begin{equation}
\label{eq:objective function f4}
%\begin{small}
\begin{aligned}
f(\boldsymbol{\tau},\mathbf{q},\mathbf{w},\boldsymbol{\varphi})\triangleq&\sum_{j \in \mathcal{J}} \sum_{k \in \mathcal{K}} \log _2 \left(1+\tau_{j, k}\right)
- \frac{1}{\ln 2}\sum_{j \in \mathcal{J}} \sum_{k \in \mathcal{K}} \tau_{j, k}\\
&+ \frac{1}{\ln 2}\bigg( \sum_{j \in \mathcal{J}} \sum_{k \in \mathcal{K}} 2 \sqrt{1+\tau_{j, k}} \mathfrak{Re}\{q_{j, k}^{*} \widetilde{\mathbf{h}}_{j,k}^{H} \mathbf{w}_{j, k}\}
-\sum_{j \in \mathcal{J}} \sum_{k \in \mathcal{K}}\left|q_{j, k}\right|^{2}\Big(\sum_{i \in \mathcal{K}}\left|\widetilde{\mathbf{h}}_{j,k}^{H} \mathbf{w}_{j, i}\right|^{2}+\sigma^{2}_k\Big)\bigg).
\end{aligned}
%\end{small}
\end{equation}
\hrulefill
\end{figure*}
which is concave with respect to each variable. %, which is tractable.
To handle the highly coupled optimization variables $\boldsymbol{\tau}$, $\mathbf{q}$, $\mathbf{w}$, and $\boldsymbol{\varphi}$ in this objective function, we adopt the BCD method to alternately update them as follows. %, as illustrated in Fig. \ref{fig:iter}, in which $t-1$ denotes the previous iteration.
%To be specific, we decompose the problem into three parts, i.e., the updating of auxiliary variables, the BS-user association and active beamforming optimization sub-problem and the BS-RIS association and passive beamforming optimization sub-problem. The algorithms we proposed to solve these problems are presented in the following subsections.
%\begin{figure}[!t]_{j,k}_{j}
%	\centering
%	\includegraphics[width = 9 cm]{iter.eps}
%	\caption{BCD method for solving problem \eqref{eq:problem_original_zeronorm}.}
%	\label{fig:iter}
%\end{figure}
\subsection{Update Auxiliary Variables $\boldsymbol{\tau}$ and $\mathbf{q}$}
With fixed $\mathbf{q}$, $\mathbf{w}$, and $\boldsymbol{\varphi}$, it is obvious that the objective function \eqref{eq:objective function f4} is concave with respect to $\tau_{j,k}$. %_{j,k} _{j}
Thus, we can easily obtain the optimal solution of $\tau_{j,k}$ by setting $\partial f / \partial \tau_{j,k}=0$ with respect to each $j \in \mathcal{J},~ k \in \mathcal{K}$. The optimal solution of $\tau_{j,k}$ is then calculated as
\begin{equation}
\label{eq:lamda}
\tau_{j,k}^{\star}  = \frac{|\widetilde{\mathbf{h}}_{j,k}^{H} \mathbf{w}_{j, k}|^{2}}{\sum\limits_{i \in \mathcal{K}, i \neq k}|\widetilde{\mathbf{h}}_{j,k}^{H} \mathbf{w}_{j, i}|^{2}+\sigma^{2}_k} , ~ \forall j, ~\forall k .
\end{equation}
%Then, by removing the irrelevant terms, the objective function \eqref{eq:objective function f4} can be transformed into
%\begin{equation}
%\label{eq:objective function f5}
%\begin{aligned}
%f_{5}(\mathbf{q})=&\sum_{j \in \mathcal{J}} \sum_{k \in \mathcal{K}} 2 \sqrt{1+\tau_{j, k}} \mathfrak{Re}\left\{q_{j, k}^{*} \widetilde{\mathbf{h}}_{j,k}^{H} \mathbf{w}_{j, k}\right\}
%-\sum_{j \in \mathcal{J}} \sum_{k \in \mathcal{K}}\big|q_{j, k}\big|^{2}\bigg(\sum_{i \in \mathcal{K}}\big|\widetilde{\mathbf{h}}_{j,k}^{H} \mathbf{w}_{j, i}\big|^{2}+\sigma^{2}_k\bigg),
%\end{aligned}
%\end{equation}
%which is concave with respect to $q_{j,k}$.
Similarly, the optimal solution of $q_{j,k}$ can be obtained by setting $\partial f / \partial q_{j,k}=0$ as
\begin{equation}
\label{eq:q}
q_{j,k}^{\star}=\frac{\sqrt{1+\tau_{j ,k}}\hspace{0.2cm} \widetilde{\mathbf{h}}_{j,k}^{H} \mathbf{w}_{j, k}}{\sum \limits_{i \in \mathcal{K}}\big|\widetilde{\mathbf{h}}_{j,k}^{H} \mathbf{w}_{j, i}\big|^{2}+\sigma^{2}_k} , ~\forall j,~ \forall k .
\end{equation}

%\subsection{BS-user Association and Active Beamforming Optimization}
\subsection{Update Active Beamforming $\mathbf{w}$}%_{j,k}
In this subsection, we introduce an MM based algorithm to optimize the active beamforming. Since the BS-user association can be represented by the active beamforming, we can easily obtain the association results based on the resulting active beamforming.
With fixed $\boldsymbol{\tau}$, $\mathbf{q}$, and $\boldsymbol{\varphi}$, the active beamforming optimization problem is reduced to %_{j}
\begin{subequations}
\begin{align}
\hspace{-0.2cm}\mathop{\max}\limits_{\substack{\mathbf{w}}} \; &\sum_{j \in \mathcal{J}} \sum_{k \in \mathcal{K}} (-\mathbf{w}_{j,k}^H \boldsymbol{\Upsilon}_j \mathbf{w}_{j,k} + \mathfrak{Re} \{ \boldsymbol{\alpha}_{j,k}^H\mathbf{w}_{j,k} \} - \varsigma_{j,k})\label{pp:obje_func}\\
\text{s.t.} \; & \sum_{k \in \mathcal{K}} \left\|\mathbf{w}_{j,k}\right\|_2^{2} \le P_{j}, ~ \forall j , \label{pp:power}\\
& \sum_{j \in \mathcal{J}}\left\|\left\|\mathbf{w}_{j, k}\right\|_2^2\right\|_0 = 1, ~ \forall k ,\label{eq:constraint_w_original}
\end{align}
\end{subequations}%_{j,k}\\ \forall j \in \mathcal{J}, \forall k \in \mathcal{K}
in which we define
\begin{subequations}
\begin{align}
\boldsymbol{\Upsilon}_j &\triangleq \sum_{k \in \mathcal{K}}|q_{j,k}|^2\widetilde{\mathbf{h}}_{j,k}\widetilde{\mathbf{h}}_{j,k}^{H},\\
\boldsymbol{\alpha}_{j,k}&\triangleq (2\sqrt{1+\tau_{j,k}}q_{j,k}^*\widetilde{\mathbf{h}}_{j,k}^{H})^H,\\
\varsigma_{j,k}&\triangleq|q_{j,k}|^2 \sigma_k^2,%\sum_{j \in \mathcal{J}} \sum_{k \in \mathcal{K}}
\end{align}
\end{subequations}
to re-arrange the objective function into a more compact form.
We clearly see that the objective function \eqref{pp:obje_func} is concave with respect to $\mathbf{w}_{j,k}$, constraint \eqref{pp:power} is convex. Thus, the non-convex constraint \eqref{eq:constraint_w_original} is the major difficulty that requires sophisticated derivations as follows.

Notice that constraint \eqref{eq:constraint_w_original} is a non-smooth non-convex constraint due to the property of $l_0$-norm.
To make this constraint \eqref{eq:constraint_w_original} more tractable, we approximate the $l_0$-norm, i.e., $\|x\|_0$, with a continuous smooth and concave function \cite{M. Tao}
\begin{equation}
\label{eq:f_delta}
g_{\delta}(x) \triangleq 1-\exp\left(-\frac{x}{\delta}\right),
\end{equation}
where $\delta > 0$ is a parameter that controls the smoothness of the approximation.
With the decreasing of $\delta$, function \eqref{eq:f_delta} becomes more approximate to $l_0$-norm with worse smoothness.
By utilizing the $l_0$-norm approximation method, constraint \eqref{eq:constraint_w_original} can be relaxed as
\begin{equation}
\label{eq:w_relax}
n_{1} \overset{\text{(c)}}\le \sum_{j \in \mathcal{J}}g_{\delta}(\|\mathbf{w}_{j, k}\|_2^2) \overset{\text{(d)}}\le n_{2}, ~ \forall k ,
\end{equation}
where
\begin{equation}
g_{\delta}(\|\mathbf{w}_{j, k}\|_2^2) = 1-\exp\big(-\frac{\|\mathbf{w}_{j, k}\|_2^2}{\delta}\big).
\end{equation}
In \eqref{eq:w_relax}, $n_{1} \in \mathbb{R}$ and $n_{2} \in \mathbb{R}$ are numbers slightly less and greater than one, respectively.
Since $g_{\delta}(\|\mathbf{w}_{j, k}\|_2^2)$ is a non-concave function with respect to $\mathbf{w}_{j, k}$, the inequalities (c) and (d) in \eqref{eq:w_relax} cannot be solved directly.
Thus, we use the MM method to deal with these two intractable constraints.
In brief, the essence of MM method is to construct an upper-bound surrogate function that locally approximates the objective function in each iteration, in order to make the optimization problem more tractable \cite{Y. Sun}.
The details of constructing surrogate functions for inequalities (c) and (d) are presented as follows.

\textbf{Inequality (c):}
To better handle the inequality (c) in \eqref{eq:w_relax}, we first convert it into a more standard form in convex optimization as
\begin{equation}
\label{eq:w_more}
-\sum_{j \in \mathcal{J}}g_{\delta}(\|\mathbf{w}_{j, k}\|_2^2) \le -n_{1},  ~ \forall k .
\end{equation}
%which is more standard in convex optimization.
%Then, we attempt to construct a convex upper-bound function of $-\sum_{j \in \mathcal{J}}h_1 \left(\mathbf{w}_{j, k}\right)$, which can be utilized as a surrogate function.
%change the non-convex constraint \eqref{eq:w_more} to a convex form.
%The key step is constructing an upper-bound of $-\sum_{j \in \mathcal{J}}h_1 \left(\mathbf{w}_{j, k}\right)$, which needs to be a convex function.
%Notice that, if we find a upper-bound function $g_1\left(\mathbf{w}_{j, k}\right)$ for each $-h_1 \left(\mathbf{w}_{j, k}\right)$, i.e.,
%\begin{equation}
%\label{eq:w_mm_second_taylor}
%\begin{aligned}
%-h_1 \left(\mathbf{w}_{j, k}\right) \le g_1\left(\mathbf{w}_{j, k}\right),
%\le -n_{\text{l}}
%~ \forall j \in \mathcal{J},   ~ \forall k \in \mathcal{K},
%\end{aligned}
%\end{equation}
%then the constraint
%\begin{equation}
%\begin{aligned}
%-\sum_{j \in \mathcal{J}}h_1 \left(\mathbf{w}_{j, k}\right) \le \sum_{j \in \mathcal{J}}g_1\left(\mathbf{w}_{j, k}\right),  ~ \forall k \in \mathcal{K} ,
%\end{aligned}
%\end{equation}
%can be satisfied.
Then, by using the second-order Taylor expansion, a convex upper-bound $\widetilde{g}_{\delta}(\mathbf{w}_{j, k}|\mathbf{w}_{j, k}^{(t-1)})$ of $-g_{\delta}(\|\mathbf{w}_{j, k}\|_2^2)$ can be constructed as shown in the following proposition, whose proof is provided in the Appendix A. %$g_1\left(\mathbf{w}_{j, k}\right)$

\noindent\textbf{Proposition 1:} The convex upper-bound $\widetilde{g}_{\delta}(\mathbf{w}_{j, k}|\mathbf{w}_{j, k}^{(t-1)})$ of $-g_{\delta}(\|\mathbf{w}_{j, k}\|_2^2)$ can be constructed as %,~ \forall j \in \mathcal{J}, ~ \forall k \in \mathcal{K},
\begin{equation}
\label{eq:g1}
\begin{aligned}
\widetilde{g}_{\delta}(\mathbf{w}_{j, k}|\mathbf{w}_{j, k}^{(t-1)}) = &\mathfrak{Re} \{\mathbf{w}_{j,k}^H \mathbf{M}_{j,k} \mathbf{w}_{j,k} \} + \mathfrak{Re} \{ \boldsymbol{\iota}_{j,k}^H \mathbf{w}_{j,k}\}\\
&+ \varpi_{j,k},  ~ \forall j ,  ~ \forall k ,
\end{aligned}
\end{equation}
in which
\begin{subequations}
\label{var1}
%\begin{small}
\begin{align}
\mathbf{M}_{j,k}  \triangleq &\max{\{0,\lambda_{j,k}\}} \cdot \mathbf{I}_M,  ~ \forall j,  \forall k,\\
\beta_{j,k}^{(t-1)} \triangleq & \frac{1}{\delta}\exp \Big(-\frac{\big\|\mathbf{w}_{j, k}^{(t-1)}\big\|_{2}^{2}}{\delta}\Big),  ~ \forall j,  \forall k ,\\
\boldsymbol{\iota}_{j,k}^H  \triangleq&-\hspace{-0.1cm}2(\mathbf{w}_{j,k}^{(t-1)})^H\mathbf{M}_{j,k} \hspace{-0.1cm}- \hspace{-0.1cm} 2\beta_{j,k}^{(t-1)}(\mathbf{w}_{j,k}^{(t-1)})^H,~  \forall j , \forall k ,\\
\varpi_{j,k}  \triangleq&- h_{\delta}(\|\mathbf{w}_{j, k}^{(t-1)}\|_2^2)+\mathfrak{Re} \{ (\mathbf{w}_{j,k}^{(t-1)})^H \mathbf{M}_{j,k} \mathbf{w}_{j,k}^{(t-1)}\}  \hspace{-0.05cm}\\ \nonumber
&+ \hspace{-0.05cm}2\mathfrak{Re} \{ \beta_{j,k}^{(t-1)} (\mathbf{w}_{j,k}^{(t-1)})^H\mathbf{w}_{j,k}^{(t-1)}\}
, \forall j ,  \forall k .
\end{align}
%\end{small}
\end{subequations}In \eqref{var1}, $\mathbf{w}_{j, k}^{(t-1)}$ is the solution obtained in the previous iteration and $\lambda_{j,k}$ is the maximum eigenvalue of the complex Hessian matrix of $-g_{\delta}(\|\mathbf{w}_{j, k}\|_2^2)$.\hfill $\blacksquare$

Thus, $\sum_{j \in \mathcal{J}}\widetilde{g}_{\delta}(\mathbf{w}_{j, k}|\mathbf{w}_{j, k}^{(t-1)})$ can be used as the surrogate upper-bound function of $-\sum_{j \in \mathcal{J}}g_{\delta}(\|\mathbf{w}_{j, k}\|_2^2)$, and the inequality constraint \eqref{eq:w_more} is transformed into
%Finally, the constraint \eqref{eq:w_more} can be transformed into
\begin{equation}
\label{eq:constraint_w_zeronorm_more}
\begin{aligned}
\sum_{j \in \mathcal{J}}\widetilde{g}_{\delta}(\mathbf{w}_{j, k}|\mathbf{w}_{j, k}^{(t-1)}) \le -n_{1},  ~ \forall k,
\end{aligned}
\end{equation}%-\sum_{j \in \mathcal{J}}h_1 \left(\mathbf{w}_{j, k}\right) \le
which is convex and can be easily handled.

\textbf{Inequality (d):}
For inequality (d) in \eqref{eq:w_relax}, our goal is also finding a convex upper-bound function of $g_{\delta}(\|\mathbf{w}_{j, k}\|_2^2)$ to convert it into a convex constraint.
It is obvious that function $g_{\delta}(\cdot)$ is strictly monotone increasing.
Thus, by introducing auxiliary variable $\mathbf{u}\triangleq [u_{1,1},...,u_{j,k},...,u_{J,K}]^T$ which satisfies
\begin{equation}
\label{eq:constraint_u}
\left\|\mathbf{w}_{j, k}\right\|_2^2-u_{j,k} \le 0 ,  ~ \forall j \in \mathcal{J},  ~ \forall k \in \mathcal{K} ,
\end{equation}
we can easily obtain $g_{\delta}(u_{j,k})$ as an upper-bound of $g_{\delta}(\|\mathbf{w}_{j, k}\|_2^2)$, i.e., the following inequality
\begin{equation}
\begin{aligned}
\hspace{-0.2cm}g_{\delta}(\|\mathbf{w}_{j, k}\|_2^2)
\le g_{\delta}\left(u_{j,k}\right) = 1-\exp\left(-\frac{u_{j,k}}{\delta}\right),  ~ \forall j,  ~ \forall k ,
\end{aligned}
\end{equation}
is satisfied. Then, we can utilize the surrogate function $g_{\delta}\left(u_{j,k}\right)$ in the inequality (d) as
\begin{equation}
\label{eq:constraint_u_zeronorm}
\sum_{j \in \mathcal{J}}g_{\delta}(u_{j,k}) \le n_{2},  ~ \forall k.
\end{equation}
Since \eqref{eq:constraint_u_zeronorm} is still a non-convex constraint,
the first-order Taylor expansion is adopted to construct a convex upper-bound function of $g_{\delta}\left(u_{j,k}\right) $ as
\begin{equation}
\begin{aligned}
%h_{\delta}\left(u_{j,k}\right)\le
\widehat{g}_{\delta}(u_{j, k}|u_{j, k}^{(t-1)})=
&\nabla g_{\delta}\big(u_{j,k}^{(t-1)}\big) \big(u_{j,k}-u_{j,k}^{(t-1)}\big)\\
&+g_{\delta}\big(u_{j,k}^{(t-1)}\big)
 ,~ \forall j,~ \forall k, \label{eq:w_mm_first_taylor}
\end{aligned}
\end{equation}
where $u_{j,k}^{(t-1)}$ is the result obtained in the previous iteration,
and the gradient $\nabla g_{\delta}\left(u_{j,k}\right)$ is obtained by %in \eqref{eq:w_mm_first_taylor}
\begin{equation}
\begin{aligned}
\nabla g_{\delta}\left(u_{j,k}\right) = \frac{1}{\delta} \exp\Big(-\frac{u_{j,k}}{\delta}\Big) ,~ \forall j ,~ \forall k  .
\end{aligned}
\end{equation}
Then, by using the convex surrogate function $\widehat{g}_{\delta}(u_{j, k}|u_{j, k}^{(t-1)})$, inequality (d) in \eqref{eq:w_relax} can be converted into
\begin{equation}
\begin{aligned}
&\sum_{j \in \mathcal{J}}\widehat{g}_{\delta}(u_{j, k}|u_{j, k}^{(t-1)})
%\nabla h_{\delta}\big(u_{j,k}^{(t-1)}\big) \big(u_{j,k}-u_{j,k}^{(t-1)}\big)
%+\sum_{j \in \mathcal{J}}h_{\delta}\big(u_{j,k}^{(t-1)}\big)
 \le n_{2},~ \forall k .
\end{aligned}
\end{equation}

In summary, based on the above derivations, the active beamforming design problem at the $t$-th iteration can be reformulated as
\begin{subequations}
\label{eq:problem_active_cvx}
\begin{align}
\mathop{\max}\limits_{\substack{\mathbf{w},\mathbf{u}}} \quad&\sum_{j \in \mathcal{J}} \sum_{k \in \mathcal{K}} (-\mathbf{w}_{j,k}^H \boldsymbol{\Upsilon}_j \mathbf{w}_{j,k} + \mathfrak{Re} \{ \boldsymbol{\alpha}_{j,k}^H\mathbf{w}_{j,k} \})\\
\text {s.t.} \quad & \sum_{k \in \mathcal{K}} u_{j,k} \le P_{j}, ~ \forall j , \\
&\left\|\mathbf{w}_{j, k}\right\|_2^2-u_{j,k} \le 0 ,  ~ \forall j,  ~ \forall k ,\\
&\sum_{j \in \mathcal{J}}\widetilde{g}_{\delta}(\mathbf{w}_{j, k}|\mathbf{w}_{j, k}^{(t-1)}) \le -n_{1},  ~ \forall k,\label{w_constraint_1}\\
&%\sum_{j \in \mathcal{J}}\nabla h_{\delta}\big(u_{j,k}^{(t-1)}\big) \big(u_{j,k}-u_{j,k}^{(t-1)}\big) +\sum_{j \in \mathcal{J}}h_{\delta}\big(u_{j,k}^{(t-1)}\big)
 \sum_{j \in \mathcal{J}}\widehat{g}_{\delta}(u_{j, k}|u_{j, k}^{(t-1)}) \le n_{2}, ~ \forall k ,\label{w_constraint_3}
\end{align}
\end{subequations}%_{j,k},u_{j,k}\\ \forall j \in \mathcal{J}, \forall k \in \mathcal{K}
which is convex and can be solved by many existing methods or convex optimization solvers, such as CVX \cite{M. Grant}.
%\subsection{BS-RIS Association and Passive Beamforming Optimization}
\subsection{Update Passive Beamforming $\boldsymbol{\varphi}$}%_j
In this subsection, we present the algorithm for optimizing the passive beamforming of RIS, which also indicates the BS-RIS association.
With fixed $\boldsymbol{\tau}$, $\mathbf{q}$, and $\mathbf{w}$, the reformulated objective function \eqref{eq:objective function f4} can be concisely re-arranged as the following form with respect to $\boldsymbol{\varphi}_{j}$%the passive beamforming optimization problem can be rewritten as_{j,k}
\begin{equation}
\label{eq:objective function f8}
\begin{aligned}
\sum_{j \in \mathcal{J}} (-\boldsymbol{\varphi}_{j}^H\mathbf{D}_j \boldsymbol{\varphi}_{j}+2\mathfrak{Re}\{\boldsymbol{\varphi}_{j}^H\mathbf{v}_j\} + c_j),
\end{aligned}
\end{equation}%f_{7}(\boldsymbol{\varphi}_{j})=&
in which%where $\mathbf{D}_j$, $\mathbf{v}_j$, and $c_j$ are defined as %in \eqref{eq:D}, \eqref{eq:v}, and \eqref{eq:c}, respectively, as shown at the top of this page.
%\begin{figure*}[!t]
\begin{subequations}
%\begin{small}
\begin{align}
\mathbf{a}_{j,i,k}\triangleq&(\operatorname{diag}\{\mathbf{h}_{\text{r}, k}^{H}\} \mathbf{G}_{j} \mathbf{w}_{j,i})^*,
~ b_{j,i,k}\triangleq(\mathbf{h}_{\text{d}, j, k}^{H} \mathbf{w}_{j,i})^*,\\
\mathbf{D}_j\triangleq&\hspace{-0.1cm}\sum_{k\in \mathcal{K}}\left|q_{j,k}\right|^{2} \sum_{i\in \mathcal{K}} \mathbf{a}_{j,i, k} \mathbf{a}_{j,i, k}^{H}, \label{eq:D}\\
\mathbf{v}_j\triangleq&\hspace{-0.1cm}\sum_{k\in \mathcal{K}}(\sqrt{1+\tau_{j,k}} q_{j,k}^{*} \mathbf{a}_{j,k, k}\hspace{-0.1cm}-\hspace{-0.1cm}\left|q_{j,k}\right|^{2} \sum_{i\in \mathcal{K}} b_{j,i, k}^{*} \mathbf{a}_{j,i, k}), \label{eq:v}\\
c_j\triangleq&\hspace{-0.1cm}\sum_{k\in \mathcal{K}}2 \sqrt{1+\tau_{j,k}} \mathfrak{Re}\left\{q_{j,k}^{*} b_{j,k, k}\right\}  \\ \nonumber
&-\sum_{k\in \mathcal{K}}\left|q_{j,k}\right|^{2}(\sigma^{2}_k+\sum_{i\in \mathcal{K}}\left|b_{j,i, k}\right|^{2}).
\end{align}
%\end{small}
\end{subequations}
%\hrulefill
%\end{figure*}
Thus, by removing the irrelevant constant term $c_j$, the passive beamforming optimization problem can be formulated as
\begin{subequations}
\label{eq:problem_f8}
\begin{align}
\hspace{-0.1cm}\mathop{\max}\limits_{\substack{\boldsymbol{\varphi}}} ~ &\sum_{j \in \mathcal{J}} (-\boldsymbol{\varphi}_{j}^H\mathbf{D}_j \boldsymbol{\varphi}_{j}+2\mathfrak{Re}\{\boldsymbol{\varphi}_{j}^H\mathbf{v}_j\})\label{eq:objective function f8}\\
\text {s.t.}~ & |\varphi_{j,n}|=1, ~ \forall j ,~ \forall n, \label{eq:unit}\\	
& \sum_{j \in \mathcal{J}}\left\|\left\|\ln(\boldsymbol{\varphi}_{j})\right\|_2^2\right\|_0 = 1,
\end{align}
\end{subequations}%_{j}\forall j \in \mathcal{J}
which is still difficult to solve due to the non-convex unit modulus constraints and non-smooth non-convex $l_0$-norm constraints. Here, the ADMM method \cite{S. Boyd} is adopted to solve this problem.
We first deal with the unit modulus constraints \eqref{eq:unit} by introducing auxiliary variables $\boldsymbol{\psi}_j \triangleq [{\psi}_{j,1},...,{\psi}_{j,N}]^T \in \mathbb{C}^{N \times 1},\forall j$, and reformulate problem \eqref{eq:problem_f8} as
\begin{subequations}
\label{eq:problem_f8_2}
\begin{align}
\mathop{\max}\limits_{\substack{\boldsymbol{\varphi}_{j},\boldsymbol{\psi}_j, \forall j }} \; &\sum_{j \in \mathcal{J}} (-\boldsymbol{\varphi}_{j}^H\mathbf{D}_j \boldsymbol{\varphi}_{j}+2\mathfrak{Re}\{\boldsymbol{\varphi}_{j}^H\mathbf{v}_j\})\\
\text {s.t.} \quad & |\varphi_{j,n}| \le 1, ~ \forall j ,~ \forall n , \\	
& |\psi_{j,n}|=1, ~ \forall j ,~ \forall n, \label{eq:constraint_psi_one}\\	
& \boldsymbol{\varphi}_j = \boldsymbol{\psi}_j,~ \forall j,\label{eq:coonstraint_psiequphi}\\
& \sum_{j \in \mathcal{J}}\left\|\big\|\ln(\boldsymbol{\varphi}_{j})\big\|_2^2\right\|_0 = 1, \label{eq:constraint_ln}
\end{align}
\end{subequations}
whose optimal solution can be obtained by solving its augmented Lagrangian (AL) problem.
Thus, by introducing dual variables $\boldsymbol{\xi}_j \triangleq [{\xi}_{j,1},...,{\xi}_{j,N}] \in \mathbb{C}^{N \times 1},~\forall j $ and penalty coefficient $\rho>0$, the AL problem of \eqref{eq:problem_f8_2} is written as
\begin{subequations}
\label{eq:problem_f8_3}
\begin{align}
\mathop{\max}\limits_{\substack{\boldsymbol{\varphi}_{j},\boldsymbol{\psi}_j,\boldsymbol{\xi}_j,\forall j }} \quad &\sum_{j \in \mathcal{J}} (-\boldsymbol{\varphi}_{j}^H\mathbf{D}_j \boldsymbol{\varphi}_{j} +2\mathfrak{Re} \{\boldsymbol{\varphi}_{j}^H\mathbf{v}_j\}\\ \nonumber
&-\mathfrak{Re} \{\boldsymbol{\xi}_j^H(\boldsymbol{\varphi}_j-\boldsymbol{\psi}_j )\}-\frac{\rho}{2 }\|\boldsymbol{\varphi}_j-\boldsymbol{\psi}_j \|_2^2)\\ \label{30a}
\text {s.t.} \quad & |\varphi_{j,n}| \le 1, ~ \forall j ,~ \forall n , \\	
& |\psi_{j,n}|=1, ~ \forall j ,~ \forall n , \\	
& \sum_{j \in \mathcal{J}}\left\|\big\|\ln(\boldsymbol{\varphi}_{j})\big\|_2^2\right\|_0 = 1,
\end{align}
\end{subequations}
which is more tractable than \eqref{eq:problem_f8_2} after removing the equality constraint \eqref{eq:coonstraint_psiequphi}
%where
%\begin{equation}
%\label{eq:objective function f10}
%\begin{aligned}
%f_{8}(\boldsymbol{\varphi}_{j})=&\sum_{j \in \mathcal{J}} (-\boldsymbol{\varphi}_{j}^H\mathbf{D}_j \boldsymbol{\varphi}_{j} +2\mathfrak{Re} \{\boldsymbol{\varphi}_{j}^H\mathbf{v}_j\})
%-\sum_{j \in \mathcal{J}}\mathfrak{Re} \{\boldsymbol{\xi}_j^H(\boldsymbol{\varphi}_j-\boldsymbol{\psi}_j )\}-\sum_{j \in \mathcal{J}}\frac{\rho}{2 }\big\|\boldsymbol{\varphi}_j-\boldsymbol{\psi}_j \big\|_2^2,
%\end{aligned}
%\end{equation}
%and the penalty coefficient $\rho$ is a positive real number.
and can be solved by alternately updating $\boldsymbol{\varphi}_{j}$, $\boldsymbol{\psi}_j$, and $\boldsymbol{\xi}_j$, as introduced in the following.
\subsubsection{Update $\boldsymbol{\varphi}_j$}
With fixed $\boldsymbol{\psi}_j$ and $\boldsymbol{\xi}_j$, the problem of optimizing $\boldsymbol{\varphi}_j$ can be written as
\begin{subequations}
\begin{align}
\mathop{\max}\limits_{\substack{\boldsymbol{\varphi}_{j},\forall j }} \quad &\sum_{j \in \mathcal{J}} (-\boldsymbol{\varphi}_{j}^H\mathbf{D}_j \boldsymbol{\varphi}_{j} +2\mathfrak{Re} \{\boldsymbol{\varphi}_{j}^H\mathbf{v}_j\}\\ \nonumber
&~~~~-\mathfrak{Re} \{\boldsymbol{\xi}_j^H(\boldsymbol{\varphi}_j-\boldsymbol{\psi}_j )\}-\frac{\rho}{2 }\|\boldsymbol{\varphi}_j-\boldsymbol{\psi}_j \|_2^2)\\
\text { s.t. } \quad & |\varphi_{j,n}| \le 1, ~ \forall j ,~ \forall n, \label{eq:constraint_phi_le1}\\	
& \sum_{j \in \mathcal{J}}\left\|\big\|\ln(\boldsymbol{\varphi}_{j})\big\|_2^2\right\|_0 = 1.\label{eq:constraint_zeronorm_ln}
\end{align}
\end{subequations}
It is obvious that the main difficulty to solve this problem is how to deal with the non-smooth non-convex $l_0$-norm constraint.
Similar to the active beamforming optimization algorithm introduced in the previous subsection, we utilize the $l_0$-norm approximation and MM based method to solve this problem.
By adopting the same concave function $g_{\delta}(\cdot)$ to approximate the non-smooth $l_0$-norm, constraint \eqref{eq:constraint_zeronorm_ln} can be relaxed as
\begin{equation}
\label{eq:ln_inequality}
n_{3} \overset{\text{(e)}}\le \sum_{j \in \mathcal{J}}g_{\delta}(\|\ln(\boldsymbol{\varphi}_{j})\|_2^2) \overset{\text{(f)}}\le n_{4},
\end{equation}
in which
\begin{equation}
g_{\delta}(\|\ln(\boldsymbol{\varphi}_{j})\|_2^2) = 1-\exp\Big(-\frac{\|\ln(\boldsymbol{\varphi}_{j})\|_2^2}{\delta}\Big),
\end{equation}
and $n_{3} \in \mathbb{R}$ and $n_{4} \in \mathbb{R}$ are numbers slightly less and greater than one, respectively.
%Then, since there exists the non-convex $\|\text{ln}(\boldsymbol{\varphi}_{j})\|_2^2$ form in the inequalities, different from the method we utilized to optimize active beamforming, we use the second-order Taylor expansion to deal with both inequality (e) and (f) in the following parts.
Then, we adopt the MM method to convert these two non-convex constraints, i.e., inequalities (e) and (f) in \eqref{eq:ln_inequality}, into more tractable forms.

\textbf{Inequality (e):} We first transform inequality (e) in \eqref{eq:ln_inequality} into a more standard form as
\begin{equation}
\label{eq:inequality_ln_b}
-\sum_{j \in \mathcal{J}}g_{\delta}(\|\ln(\boldsymbol{\varphi}_{j})\|_2^2) \le -n_{3}.
\end{equation}
The convex upper-bound function $\overline{g}_{\delta}(\boldsymbol{\varphi}_j|\boldsymbol{\varphi}_j^{(t-1)})$ of $-g_{\delta}(\|\ln(\boldsymbol{\varphi}_{j})\|_2^2)$ can be constructed by second-order Taylor expansion as shown in the following
proposition, whose proof is presented in Appendix B.

\noindent\textbf{Proposition 2:}
The convex upper-bound $\overline{g}_{\delta}(\boldsymbol{\varphi}_j|\boldsymbol{\varphi}_j^{(t-1)})$ of $-g_{\delta}(\|\ln(\boldsymbol{\varphi}_{j})\|_2^2)$ is written as
\begin{equation}
\label{eq:g2_upper}
\begin{aligned}
\hspace{-0.3cm}\overline{g}_{\delta}(\boldsymbol{\varphi}_j|\boldsymbol{\varphi}_j^{(t-1)}) = \mathfrak{Re} \{\boldsymbol{\varphi}_j^H \boldsymbol{\Xi}_{j} \boldsymbol{\varphi}_j \} \hspace{-0.05cm}+ \hspace{-0.05cm}\mathfrak{Re} \{ \boldsymbol{\kappa}_{j}^H\boldsymbol{\varphi}_j\}\hspace{-0.05cm} +\hspace{-0.05cm} \mu_{j},\forall j,
\end{aligned}
\end{equation}
in which
\begin{subequations}
\label{eq:h2h2}
\begin{align}
\hspace{-1cm}\mathbf{\Xi}_j \triangleq& \max{\{0,\lambda_{j}\}} \cdot \mathbf{I}_N, ~ \forall j ,\\
\boldsymbol{\kappa}_{j} \triangleq&-2(\boldsymbol{\varphi}_j^{(t-1)})^H\boldsymbol{\Xi}_{j} - 2\varepsilon_j^{(t-1)} (\boldsymbol{\phi}_j^{(t-1)})^H ,  ~ \forall j ,  \\
\mu_{j} \triangleq&- g_{\delta}(\|\ln(\boldsymbol{\varphi}_{j}^{(t-1)})\|_2^2)\hspace{-0.05cm}+\hspace{-0.05cm}\mathfrak{Re} \{ (\boldsymbol{\varphi}_{j}^{(t-1)})^H \boldsymbol{\Xi}_{j} \boldsymbol{\varphi}_{j}^{(t-1)}\}\hspace{-0.05cm} \\ \nonumber
&+ \hspace{-0.05cm} 2\mathfrak{Re} \{\varepsilon_j^{(t-1)} (\boldsymbol{\phi}_j^{(t-1)})^H \boldsymbol{\varphi}_{j}^{(t-1)}  \}
, ~ \forall j .
\end{align}
\end{subequations}
Here, for brevity we define
\begin{subequations}
\begin{align}
\varepsilon_j^{(t-1)} \triangleq& \frac{1}{\delta}\exp\Big(-\frac{\big\|\ln(\boldsymbol{\varphi}_{j}^{(t-1)})\big\|_2^2}{\delta}\Big), ~ \forall j,\\
(\boldsymbol{\phi}_j^{(t-1)})^H \triangleq&\ln({\boldsymbol{\varphi}_{j}^{(t-1)})}^H \oslash {(\boldsymbol{\varphi}_{j}^{(t-1)}) }^T, ~ \forall j.
\end{align}
\end{subequations}
In \eqref{eq:h2h2}, $\boldsymbol{\varphi}_j^{(t-1)}$ and $\lambda_{j}$ are the optimal solution obtained in the previous iteration and the maximum eigenvalue of the complex Hessian matrix of $-g_{\delta}\big(\|\ln(\boldsymbol{\varphi}_{j})\|_2^2\big)$, respectively.\hfill $\blacksquare$

Then, with the convex surrogate function $\overline{g}_{\delta}(\boldsymbol{\varphi}_j|\boldsymbol{\varphi}_j^{(t-1)})$, the inequality (e) in \eqref{eq:ln_inequality} can be transformed into
\begin{equation}
\label{eq:phi_cons1_cvx}
\sum_{j \in \mathcal{J}}\overline{g}_{\delta}(\boldsymbol{\varphi}_j|\boldsymbol{\varphi}_j^{(t-1)}) \le -n_{3},
\end{equation}%-\sum_{j \in \mathcal{J}}h_2\left(\boldsymbol{\varphi}_j\right) \le
which satisfies the original constraint and is more tractable.

\textbf{Inequality (f):} The second-order Taylor expansion is utilized to construct the upper-bound function $\breve{g}_{\delta}(\boldsymbol{\varphi}_j|\boldsymbol{\varphi}_j^{(t-1)})$ for the non-convex function $g_{\delta}(\|\ln(\boldsymbol{\varphi}_{j})\|_2^2)$ as presented in the following proposition, whose proof is shown in Appendix C.

\noindent\textbf{Proposition 3:}
The convex upper-bound $\breve{g}_{\delta}(\boldsymbol{\varphi}_j|\boldsymbol{\varphi}_j^{(t-1)})$ of $g_{\delta}(\|\ln(\boldsymbol{\varphi}_{j})\|_2^2)$ is written as
\begin{equation}
\label{eq:in_d_taylor}
\begin{aligned}
\hspace{-0.3cm}\breve{g}_{\delta}(\boldsymbol{\varphi}_j|\boldsymbol{\varphi}_j^{(t-1)}) = \mathfrak{Re} \{\boldsymbol{\varphi}_j^H \boldsymbol{\Gamma}_{j} \boldsymbol{\varphi}_j \} + \mathfrak{Re} \{ \boldsymbol{\zeta}_{j}^H\boldsymbol{\varphi}_j\} + \eta_{j}, \forall j ,
\end{aligned}
\end{equation}
where
\begin{subequations}
\begin{align}
\mathbf{\Gamma}_j \triangleq& \max{\{0,\lambda_{j}\}} \cdot \mathbf{I}_N, ~ \forall j ,\\
\boldsymbol{\zeta}_{j}^H \triangleq&-2(\boldsymbol{\varphi}_j^{(t-1)})^H\boldsymbol{\Gamma}_{j} + 2\varepsilon_j^{(t-1)} (\boldsymbol{\phi}_j^{(t-1)})^H,  ~ \forall j,  \\
\eta_{j} \triangleq&g_{\delta}(\|\ln(\boldsymbol{\varphi}_{j}^{(t-1)})\|_2^2)+\mathfrak{Re} \{ (\boldsymbol{\varphi}_{j}^{(t-1)})^H \boldsymbol{\Gamma}_{j} \boldsymbol{\varphi}_{j}^{(t-1)}\}\\ \nonumber
& - 2 \mathfrak{Re} \{\varepsilon_j^{(t-1)} (\boldsymbol{\phi}_j^{(t-1)})^H \boldsymbol{\varphi}_{j}^{(t-1)}  \}
 , ~ \forall j.
\end{align}
\end{subequations}
Here, $\lambda_{j}$ is the maximum eigenvalue of the complex Hessian matrix of $g_{\delta}(\|\ln(\boldsymbol{\varphi}_{j})\|_2^2)$.\hfill $\blacksquare$
%Similar to the matrix $\mathbf{M}_{j,k}$ that we introduced in the active beamforming part, we set $\mathbf{\Gamma}_j = \varsigma_{\text{max}}\mathbf{I}_N$, where $\varsigma_{\text{max}} = \max{\{0,\lambda_{j}\}}$.
%$\lambda_{j}$ is the maximum eigenvalue of  $\mathcal{H}_{j}(h_3)$.

Thus, by utilizing the convex upper-bound surrogate function $\sum_{j \in \mathcal{J}}\breve{g}_{\delta}(\boldsymbol{\varphi}_j|\boldsymbol{\varphi}_j^{(t-1)})$, the non-convex inequality (f) in \eqref{eq:ln_inequality} can be converted into
\begin{equation}
\label{eq:phi_cons2_cvx}
\sum_{j \in \mathcal{J}}\breve{g}_{\delta}(\boldsymbol{\varphi}_j|\boldsymbol{\varphi}_j^{(t-1)}) \le n_{4}.
\end{equation}%\sum_{j \in \mathcal{J}}h_2\left(\boldsymbol{\varphi}_j\right) \le

In summary, the optimization problem of passive beamforming at the $t$-th iteration can be written as
\begin{subequations}
\label{eq:phi_cvx}
\begin{align}
\mathop{\max}\limits_{\substack{\boldsymbol{\varphi}_{j},\forall j}} \quad &\sum_{j \in \mathcal{J}} (-\boldsymbol{\varphi}_{j}^H\mathbf{D}_j \boldsymbol{\varphi}_{j} +2\mathfrak{Re} \{\boldsymbol{\varphi}_{j}^H\mathbf{v}_j\}\\ \nonumber
&~~~~-\mathfrak{Re} \{\boldsymbol{\xi}_j^H(\boldsymbol{\varphi}_j-\boldsymbol{\psi}_j )\}-\frac{\rho}{2 }\|\boldsymbol{\varphi}_j-\boldsymbol{\psi}_j \|_2^2)\\
\text {s.t.} \quad & |\varphi_{j,n}| \le 1, ~ \forall j ,~ \forall n ,\\
& \sum_{j \in \mathcal{J}}\overline{g}_{\delta}(\boldsymbol{\varphi}_j|\boldsymbol{\varphi}_j^{(t-1)}) \le -n_{3},\label{phi_constraint_1}\\
& \sum_{j \in \mathcal{J}}\breve{g}_{\delta}(\boldsymbol{\varphi}_j|\boldsymbol{\varphi}_j^{(t-1)}) \le n_{4},\label{phi_constraint_2}
\end{align}
\end{subequations}
which is a convex problem and can be easily solved by CVX.

\subsubsection{Update $\boldsymbol{\psi}_j$}
With fixed $\boldsymbol{\varphi}_{j}$ and $\boldsymbol{\xi}_j$, the problem of optimizing $\boldsymbol{\psi}_j$ can be written as
\begin{subequations}
\label{eq:psi_pro}
\begin{align}
\mathop{\max}\limits_{\substack{\boldsymbol{\psi}_j,\forall j }} \; &-\sum_{j \in \mathcal{J}}\mathfrak{Re} \{\boldsymbol{\xi}_j^H(\boldsymbol{\varphi}_j-\boldsymbol{\psi}_j )\}-\sum_{j \in \mathcal{J}}\frac{\rho}{2}\|\boldsymbol{\varphi}_j-\boldsymbol{\psi}_j\|_2^2 \label{eq:psifun}\\
\text { s.t. } \;
& |\psi_{j,n}|=1, ~ \forall j,~ \forall n.
\end{align}
\end{subequations}
Then, to facilitate algorithm development, we rewrite the objective function (43a) as
\begin{equation}
\begin{aligned}
\sum_{j \in \mathcal{J}}\mathfrak{Re} &\{ (\boldsymbol{\xi}_j+{\rho}\boldsymbol{\varphi}_j)^H \boldsymbol{\psi}_j \} -\sum_{j \in \mathcal{J}}\frac{{\rho}}{2}\|\boldsymbol{\varphi}_j \|_2^2\\
&-\sum_{j \in \mathcal{J}}\frac{{\rho}}{2}\|\boldsymbol{\psi}_j \|_2^2 - \sum_{j \in \mathcal{J}}\mathfrak{Re}\{\boldsymbol{\xi}_j^H\boldsymbol{\varphi}_j\}.
\end{aligned}
\end{equation}
%in which $\boldsymbol{\varphi}_j$ and $\boldsymbol{\xi}_j$ are fixed constants. Moreover,
Since $\psi_{j,n}$ has unit modulus, the quadratic term $\|\boldsymbol{\psi}_j \|_2^2$ equals to $N$.
Hence, by removing the constant parts, the equivalent problem of (43) can be rewritten as
\begin{subequations}
\begin{align}
\mathop{\max}\limits_{\substack{\boldsymbol{\psi}_j,\forall j }} \quad &\sum_{j \in \mathcal{J}}\mathfrak{Re} \{(\boldsymbol{\xi}_j+{\rho}\boldsymbol{\varphi}_j)^H \boldsymbol{\psi}_j\}\\
\text { s.t. } \quad
& |\psi_{j,n}|=1, ~ \forall j,~ \forall n ,
\end{align}
\end{subequations}
whose optimal solution of $\boldsymbol{\psi}_j$ can be obtained by
\begin{equation}
\label{eq:psi}
\boldsymbol{\psi}_j^\star = e^{\jmath\angle (\boldsymbol{\xi}_j+{\rho}\boldsymbol{\varphi}_j)}.
\end{equation}

\subsubsection{Update $\boldsymbol{\xi}_j$}
With fixed $\boldsymbol{\varphi}_{j}$ and $\boldsymbol{\psi}_j$, we can update the dual variable $\boldsymbol{\xi}_j$ by
\begin{equation}
\label{eq:xi}
\boldsymbol{\xi}_j := \boldsymbol{\xi}_j +\rho(\boldsymbol{\varphi}_j-\boldsymbol{\psi}_j),
\end{equation}
which is introduced in the standard ADMM method \cite{S. Boyd}.

Finally, the ADMM based passive beamforming design algorithm is summarized in Algorithm \ref{alg:Algorithm admm}. We can obtain the solution of problem \eqref{eq:problem_f8_3} in an iterative manner by alternately updating $\boldsymbol{\varphi}_j$, $\boldsymbol{\psi}_j$, and $\boldsymbol{\xi}_j$ until convergence.

\subsection{Summary}

Based on the above derivations, the proposed FP-MM-ADMM based joint association and beamforming design algorithm is summarized in Algorithm \ref{alg:Algorithm 1}. With an appropriate initialization of active beamforming $\mathbf{w}_{j,k}, \forall j , \forall k,$ and passive beamforming $\boldsymbol{\varphi}_{j}$, the dual variable $\boldsymbol{\tau}$, the auxiliary variable $\mathbf{q}$, the active beamforming $\mathbf{w}_{j,k}$, and the passive beamforming $\boldsymbol{\varphi}_j$ are alternately updated until convergence. Finally, to achieve better performance, we select the resulting BS-RIS-user association and re-design the active and passive beamforming under the determined association result. To be specific, the optimal BS-$j^{\star}$ that serves user-$k$ is determined by
%\begin{equation}
$j^{\star} = \arg \max _{\forall j }\; g_{\delta}(\|\mathbf{w}_{j, k}\|_2^2)$.
%\end{equation}
Similarly, the optimal BS-$j^{\star}$ which is assisted by the RIS is selected by
$j^{\star} = \arg \max _{\forall j }\; g_{\delta}(\|\ln(\boldsymbol{\varphi}_{j})\|_2^2)
$. %\end{equation}

\begin{algorithm}[!t]
\caption{Proposed ADMM based passive beamforming design algorithm.}
\label{alg:Algorithm admm}
\begin{algorithmic}[1]
\REQUIRE $\delta$, $\rho$, $\boldsymbol{\tau}$, $\mathbf{q}$, $\mathbf{D}_j$, and $\mathbf{v}_j$, $\forall j$.
\ENSURE  $\boldsymbol{\varphi}_{j}, \forall j $.
\STATE {Initialize $\boldsymbol{\psi}_j \text{ and } \boldsymbol{\xi}_j, \forall j $.}
\REPEAT
    \STATE{Calculate passive beamforming $\boldsymbol{\varphi}_j,\forall j $ by solving \eqref{eq:phi_cvx};}
    \STATE{Calculate auxiliary variable $\boldsymbol{\psi}_j$ by \eqref{eq:psi};}
    \STATE{Update Lagrangian multiplier $\boldsymbol{\xi}_j$ by \eqref{eq:xi};}
\UNTIL convergence.
\RETURN $\boldsymbol{\varphi}_{j}, \forall j $.
\end{algorithmic}
\end{algorithm}

\begin{algorithm}[!t]
\caption{Proposed joint association and beamforming design algorithm.}
\label{alg:Algorithm 1}
\begin{algorithmic}[1]
\REQUIRE $P_j$, $\delta$, $\rho$, $\mathbf{h}_{\text{d},j,k}$, $\mathbf{G}_{j}$, and $\mathbf{h}_{\text{r}, k}$, $\forall j ,\forall k $.
\ENSURE  BS-RIS-user association, $\boldsymbol{\varphi}\text{ and } \mathbf{w}$.
\STATE {Initialize $\boldsymbol{\varphi}_{j}\text{ and } \mathbf{w}_{j,k}, \forall j , \forall k$.}
\REPEAT
    \STATE{Calculate dual variable $\boldsymbol{\tau}$ by \eqref{eq:lamda};}
    \STATE{Calculate auxiliary variable $\mathbf{q}$ by \eqref{eq:q};}
    \STATE{Update active beamforming $\mathbf{w}_{j,k},\forall j , \forall k $ by solving \eqref{eq:problem_active_cvx};}
    %\STATE{Update $\mathbf{D}_j$ and $\mathbf{v}_j, \forall j \in \mathcal{J}, \forall k \in \mathcal{K}$ by \eqref{eq:D} and \eqref{eq:v}, respectively;}
    \STATE{Update passive beamforming $\boldsymbol{\varphi}_j,\forall j $ by Algorithm \ref{alg:Algorithm admm};}
    %\STATE{Update the channels, in which $\boldsymbol{\Phi}_{j}\triangleq\operatorname{diag}\left(\boldsymbol{\varphi}_{j}\right)$;}
\UNTIL convergence.
\STATE{Select optimal BS-RIS-user association.}
\STATE{Re-design active and passive beamforming.}
\RETURN resulting BS-RIS-user association, $\boldsymbol{\varphi}\text{ and } \mathbf{w}$.
\end{algorithmic}
\end{algorithm}

%Moreover, the MM method requires that the initial points are in the feasible set of the optimization problem. Thus, we initialize the active beamforming $\mathbf{w}_{j,k}$ by the existing ZF precoding technology under the assumption that each user is served by all BSs, i.e., from each BS-$j$ to each user-$k$, the active beamforming $\mathbf{w}_{j,k} \ne \mathbf{0}_M$. Notice that, this initialization method can satisfy the constraint \eqref{w_constraint_1}-\eqref{w_constraint_3} because we have relaxed the assume that the transmit power at each BS has a very minimal value to satisfy  in the active beamforming problem. Similarly,
For the proposed algorithm, an appropriate initialization is needed. To ensure the fairness between users and BSs, we assume that each user receives the signals from all BSs and the RIS provides tunable phase-shifts for all BSs simultaneously, i.e., the active beamforming $\mathbf{w}_{j,k} \ne \mathbf{0}_M$ and the phase-shifts $\boldsymbol{\theta}_{j}\ne \mathbf{0}_N$, the transmit power of each $\mathbf{w}_{j,k}$ is the same $P_{j,k} = P, ~ \forall j  , \forall k $, and each $\|\boldsymbol{\theta}_{j}\|^2$ has the same value.
%For, the initial points need to, in which it is obvious that $\|\mathbf{w}_{j, k}\|_2^2$ equals to the transmit power  $P_{j,k}$ from BS-$j$ to user-$k$. For fairness, we further assume that. Thus, based on derivations, we can obtain that only when $P$ within the range
%Moreover, the MM method requires a feasible initial point. However, the above assumptions do not satisfy constraints \eqref{eq:BS constraint} and  \eqref{eq:theta constraint}.
%Since we have relaxed the constraints \eqref{eq:BS constraint} and  \eqref{eq:theta constraint} into \eqref{w_constraint_1}-\eqref{w_constraint_3} and \eqref{phi_constraint_1}, \eqref{phi_constraint_2}, respectively, there exists feasible initial active beamforming $\mathbf{w}_{j,k}$ and passive beamforming $\boldsymbol{\varphi}_{j}$, which satisfy the above assumptions.
%Then, the active beamforming $\mathbf{w}_{j,k}$ is initialized by the zero-forcing (ZF) precoding method and the passive beamforming $\boldsymbol{\varphi}_{j}$ is initialized randomly. Notice that, we need to set small initial transmit power for each BS to make the initialization results in the feasible region which consists of the constraints \eqref{w_constraint_1}-\eqref{w_constraint_3}. Similarly, each phase-shift of RIS for each BS should be set small to satisfy constraints \eqref{phi_constraint_1} and \eqref{phi_constraint_2}.
Based on the principle of MM method, the initial point should lie in the feasible region, i.e., the active beamforming $\mathbf{w}_{j,k}$ should satisfy constraint \eqref{eq:w_relax}, which leads to
\begin{equation}
\delta \ln (\frac{J}{J-n_1}) \le P \le \delta \ln (\frac{J}{J-n_2}), ~ \forall k .
\end{equation}
It is obvious that a small $P$ should be guaranteed since $\delta\ll 1$, $n_{1}$ and $n_{2}$ are slightly less and greater than one.
After choosing an appropriate initial power $P$, the active beamforming $\mathbf{w}_{j,k}$ is initialized by the typical zero-forcing (ZF) precoding method for the purpose of effectively suppressing the multi-user interference.
Similarly, the initial phase-shifts of passive beamforming $\boldsymbol{\varphi}_{j}$ should be generated under the constraint \eqref{eq:ln_inequality}, which also requires small value of $\|\ln(\boldsymbol{\varphi}_{j})\|_2^2$ (i.e., $\|\boldsymbol{\theta}_{j}\|_2^2$). The range of $\boldsymbol{\theta}_{j}$ can be calculated as
\begin{equation}
\delta \ln (\frac{J}{J-n_3}) \le \|\boldsymbol{\theta}_{j}\|_2^2 \le \delta \ln (\frac{J}{J-n_4}).
\end{equation}
%Then, based on the given transmit power, we initialize the active beamforming $\mathbf{w}_{j,k}$ by.
%Similarly, the phase-shifts of passive beamforming $\boldsymbol{\varphi}_{j}$ are randomly initialized and each phase-shift of RIS for each BS should be set small to satisfy the constraint.
%the active beamforming $\mathbf{w}_{j,k}$ is obtained by the existing precoding strategy, i.e., zero-forcing (ZF) precoding method, from the feasible set under the assumption that each user is served by all BSs. Moreover, the passive beamforming $\boldsymbol{\Phi}_{j}$ is randomly generated from its feasible set.

Then, we provide an analysis of the complexity for the proposed joint association and beamforming design algorithm. In each iteration of Algorithm \ref{alg:Algorithm 1}, updating the dual variable $\boldsymbol{\tau}$ has a complexity of approximately $\mathcal{O}\{J K(K+1) M N^{2}\}$, updating auxiliary variable $\mathbf{q}$ requires about $\mathcal{O}\{J K^{2} M N^{2}\}$ operations, updating the active beamforming $\mathbf{w}_{j,k}$ by CVX has a complexity of $\mathcal{O}\{ \sqrt{(J+3K+2JK)}[J^4 K^4 M^2 (J   +  K )]\}$. In each iteration of Algorithm \ref{alg:Algorithm admm}, the major complexity comes from updating the passive beamforming $\boldsymbol{\varphi}_j$ by CVX, which has a complexity of  $\mathcal{O}\{\sqrt{2(JN+2)} J^2 N^2\}$. Therefore, the total complexity of the proposed algorithm is of order $\mathcal{O}\{I_\text{p}[ J (2K+1) M N^{2}+\sqrt{(J+3K+2JK)}(J^4 K^4 M^2 (J   +  K )) + I_\text{ADMM}\sqrt{2(JN+2)} J^2 N^2]\}$, in which $I_\text{p}$ and $I_\text{ADMM}$ are the iterations of the proposed joint association and beamforming design algorithm and the ADMM based passive beamforming design algorithm, respectively.

\section{Simulation Results}
\begin{figure}[!t]
    \centering
	\includegraphics[width = 3.0 in]{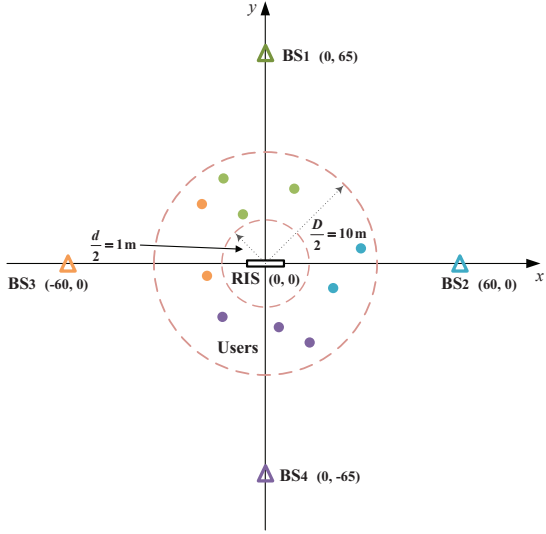}
	\caption{An illustration of the position of BSs, RIS, and users.}
	\label{fig:location}
\end{figure}

In this section, we provide extensive simulation results to demonstrate the effectiveness of our proposed FP-MM-ADMM based joint association and beamforming design algorithm for a RIS-assisted multi-cell wireless network. We consider that $J = 4$ BSs are located at $(0\text{m}, 65\text{m})$, $(60 \text{m}, 0 \text{m})$, $(-60\text{m}, 0 \text{m})$, and $(0 \text{m}, -65 \text{m})$, respectively, as shown in Fig. \ref{fig:location}.
Each BS is equipped with $M = 32$ antennas.
The RIS consisting of $N = 64$ reflecting elements is deployed at $(0 \text{m}, 0 \text{m})$ and $K = 15$ single-antenna users are randomly distributed within a ring area centered at $(0 \text{m}, 0\text{m})$ with an inner diameter of $2 \text{m}$ and an outer diameter of $20 \text{m}$. The noise power $\sigma^2_k$ at the receivers is set as $-80 \text{ dBm}$. %and the maximum tranmit power of the whole network $P_{\text{max}}$ are set as $-80 \text{ dBm}$ and $30 \text{ dBm}$, respectively.
We assume that each BS has the same transmit power, i.e., $P_j = P_{\text{max}}/J$. %The two-dimensional coordinate system is shown in Fig. \ref{fig:location}, which illustrates the position relationship of BSs, RIS, and users.
%\begin{figure}[!t]
%	\centering
%	\includegraphics[width = 9 cm]{location.eps}
%	\caption{An illustration of the position of BSs, RIS, and users.}
%	\label{fig:location}
%\end{figure}

In addition, the path loss is modeled as $L(d)=C_{0}(\frac{d}{D_{0}})^{-\alpha}$, where $d$ is the distance, $C_0$ and $D_0$ are set as $-30\text{ dB}$ and $1\text{ m}$, respectively \cite{Q. Wu3}. The path loss exponents are set as 3.5, 2.5 and 2.8 for the BS-user, BS-RIS, and RIS-user channels, respectively. Moreover, the Rician channel model is considered in the simulation studies. The BS-user channels are assumed as Non-Line-of-Sight (NLoS) channels because of the complex wireless propagation environment between BSs and users. Since RIS is usually deployed at high buildings, there exist strong Line-of-Sight (LoS) channels between BSs and RIS. Furthermore, the RIS-user channels consist both LoS links and NLoS links since the RIS is set nearby the users. Thus, we set the Rician factors as 0, $\infty$, and 1 for the BS-user, BS-RIS, and RIS-user channels, respectively.

\begin{figure}[h]
\centering
\subfigure[First inner loop.]
{\label{fig:subfig:admm}
\includegraphics[width=0.47\linewidth]{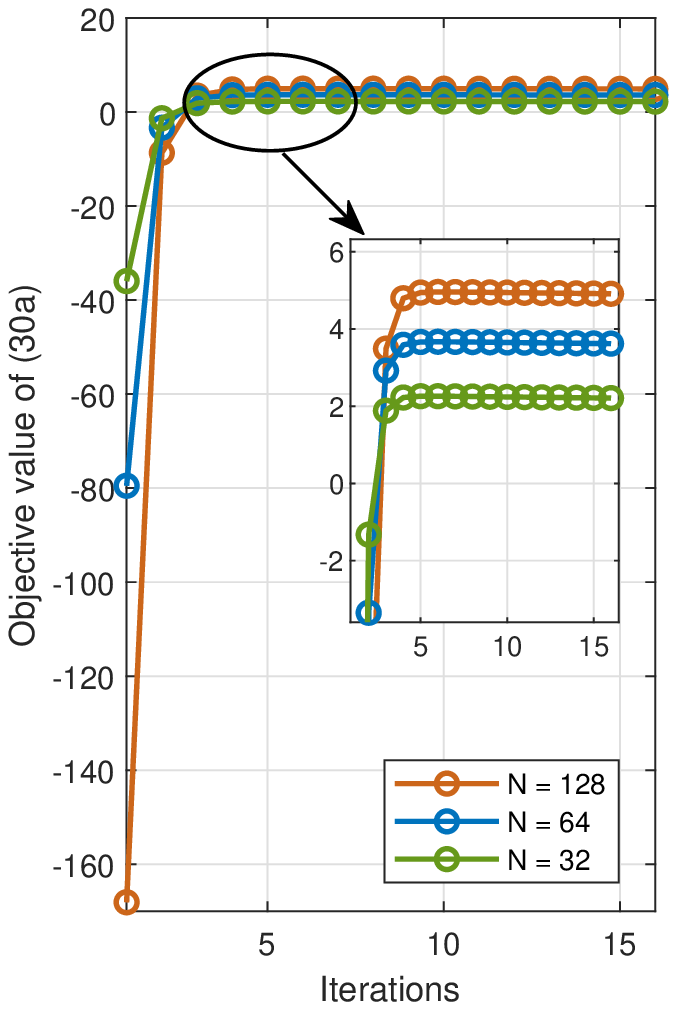}}
%\hspace{0.01\linewidth}
\subfigure[Outer loop.]
{\label{fig:subfig:outer}
\includegraphics[width=0.47\linewidth]{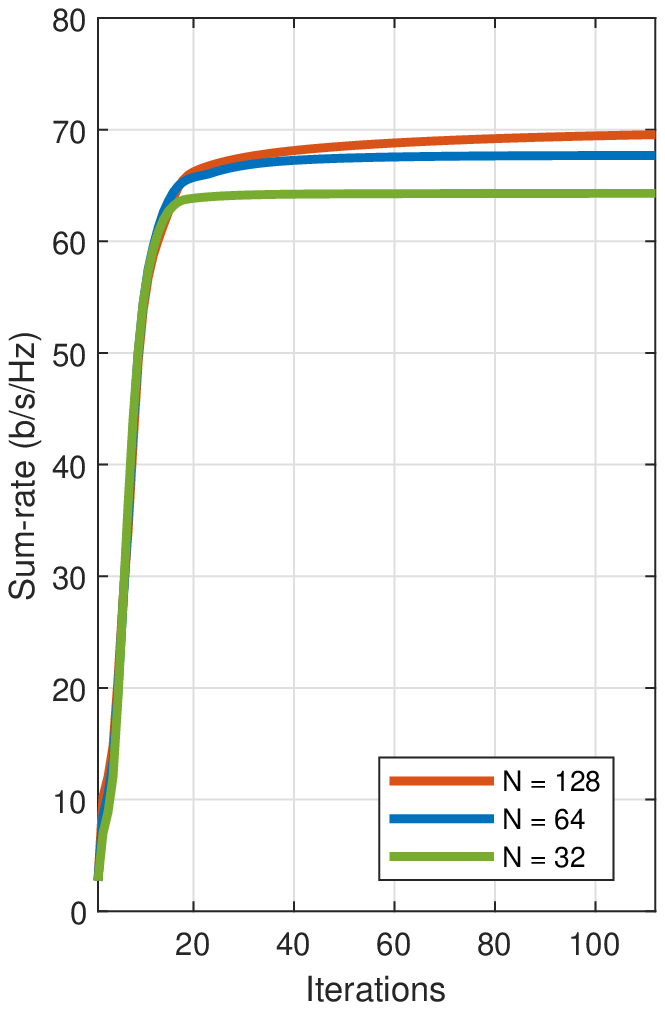}}
\caption{Convergence of proposed algorithm.}
\label{fig:convergence}
\end{figure}

%We first present the convergence of the proposed FP-MM-ADMM based joint association and beamforming design algorithm by plotting the sum-rate versus the number of iterations in Fig. \ref{fig:convergence}.
%It can be seen that the convergence can be achieved within a limited number of iterations under different settings.
%In addition, since we initialize the active beamforming under the assumption that each BS has the same minimal transmit power, more steps are required for optimizing the active beamforming to meet the maximum power constraint with the increasing of $P_{\text{max}}$, which causes slower convergence.
%\begin{figure}[!t]
%    \centering
%	\includegraphics[width = 3.3 in]{convergence.eps}
%	\caption{Convergence of joint association and beamforming design algorithm.}
%	\label{fig:convergence}
%\end{figure}
We first present the convergence performance of the proposed FP-MM-ADMM based joint association and beamforming design algorithm by plotting the objective value of (30a) and sum-rate versus the number of iterations in Fig. \ref{fig:convergence}, respectively.
To be specific, Fig. \ref{fig:subfig:admm} verifies the convergence of the first loop of Algorithm \ref{alg:Algorithm admm} and Fig. \ref{fig:subfig:outer} demonstrates the convergence of Algorithm \ref{alg:Algorithm 1}.
It can be seen that the convergence can be achieved within a limited number of iterations under different settings.
%In addition, since we initialize the active beamforming under the assumption that each BS has the same minimal transmit power, more steps are required for optimizing the active beamforming to meet the maximum power constraint with the increasing of $P_{\text{max}}$, which causes slower convergence.

\begin{figure}[!t]
	\centering
	\includegraphics[width = 3.5 in]{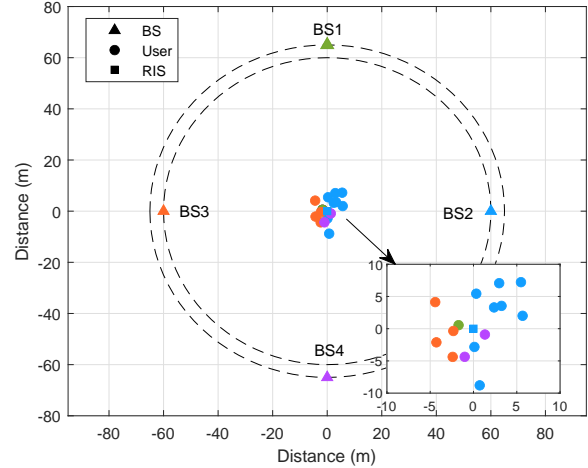}%{test1_circle.eps}
	\caption{Association results ($P_{\text{max}} = 20\text{dBm}$, $M = 32$, $N = 64$, $K = 15$).}
	\label{fig:association}
\end{figure}

%Then, the association results of our proposed algorithm is shown in Fig. \ref{fig:association}, in which the different colors represent the different BSs as well as their associated RIS and users. We can observe that BS4, i.e., the purple BS, associated with less users compared with other BSs, which may caused by its worst channel gain and the fixed phase-shift interference of RIS, which is associated the orange BS.
Then, the association results of our proposed algorithm is shown in Fig. \ref{fig:association}, in which different colors represent different BSs as well as their associated RIS and users.
We can observe that the RIS is associated to BS2, i.e., the blue BS, which can significantly enlarge the coverage area of this cell.
Thus, more users are associated to BS2 to achieve better sum-rate performance.
Moreover, it is obvious that BS1, i.e., the green BS, is associated with less users due to the worst channel gain.

\begin{figure}[!t]
    \centering
    \includegraphics[width = 3.5 in]{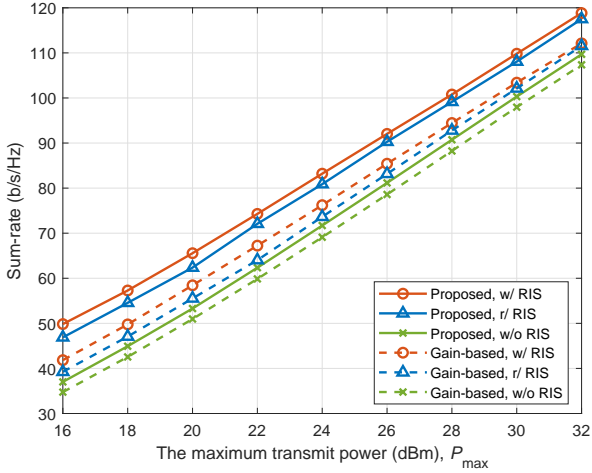}
    \caption{Sum-rate versus the maximum transmit power, $P_{\text{max}}$ ($M = 32$, $N = 64$, $K = 15$).}
    \label{fig:changeP}
\end{figure}

%\begin{figure}[!t]
%	\centering
%	\includegraphics[width = 9 cm]{admm_convergence.eps}
%	\caption{Convergence of ADMM based passive beamforming design algorithm.}
%	\label{fig:admm_convergence}
%\end{figure}
%\begin{figure}[!t]
%	\centering
%	\includegraphics[width = 9 cm]{convergence.eps}
%	\caption{Convergence of joint association and beamforming design algorithm.}
%	\label{fig:convergence}
%\end{figure}
Fig. \ref{fig:changeP} evaluates the proposed algorithm by showing the sum-rate versus the maximum transmit power $P_{\text{max}}$. For comparison purpose, we also include the traditional direct channel gain based user association scheme, which determines the BS-user association only based on the channel gain of the direct BS-user channels \cite{A. H. Sakr}, \cite{P. Ni} and optimizes the BS-RIS association by the proposed ADMM based passive beamforming design algorithm.
%Since higher channel gain indicates lower path loss, users will be associated with the BS which has the lowest BS-user path loss by utilizing the gain-based algorithm.
Since the gain-based algorithm only considers the direct BS-user channels, the resulting BS-user association will not be affected by the deployment of RIS. % which can more distinctly .
%\textcolor{blue}{This channel gain based association algorithm theoretically leads to better system performance since it can associate users with the BSs which have the minimum path loss.}\cite{D. Liu},
To better demonstrate the advantages of the proposed joint BS-RIS-user association and beamforming design, these two schemes are compared under three scenarios. i) With RIS (w/ RIS): A RIS is deployed in the network, the BS-RIS association and its passive beamforming are optimized; ii) Random RIS (r/ RIS): A RIS is deployed in the network, but the BS-RIS association and its passive beamforming are randomly generated; iii) Without RIS (w/o RIS): The case without deploying RIS.
%\begin{figure}[!t]
%	\centering
%	\includegraphics[width = 9 cm]{changeP.eps}
%	\caption{Sum-rate versus the maximum transmit power, $P_{\text{max}}$.}
%	\label{fig:changeP}
%\end{figure}
It can be observed from Fig. \ref{fig:changeP} that the ``w/ RIS" and ``r/ RIS" schemes outperform the ``w/o RIS" scheme, which demonstrates the advantage of deploying RIS in the multi-cell wireless communication network.
Moreover, the proposed algorithm can achieve better performance compared with the gain-based association algorithm for all power ranges. This is because the gain-based association algorithm only considers the BS-user channel gain and ignores the impact of the BS-RIS-user cascaded channel in the process of BS-user association, which causes significant performance loss.
Furthermore, Fig. \ref{fig:changeP} shows that the performance gap between ``w/ RIS" and ``w/o RIS" schemes for the proposed algorithm is larger than that for the gain-based algorithm, which further %means that the proposed algorithm is more applicable to the RIS-assisted multi-cell networks. These results
verifies the importance of the joint design of BS-RIS-user association and beamforming.
%\begin{figure}[!t]
%	\centering
%	\includegraphics[width = 9 cm]{changeM.eps}
%	\caption{Sum-rate versus the number of BS antennas, $M$.}
%	\label{fig:changeM}
%\end{figure}
\begin{figure}[!t]
    \centering
    \includegraphics[width = 3.5 in]{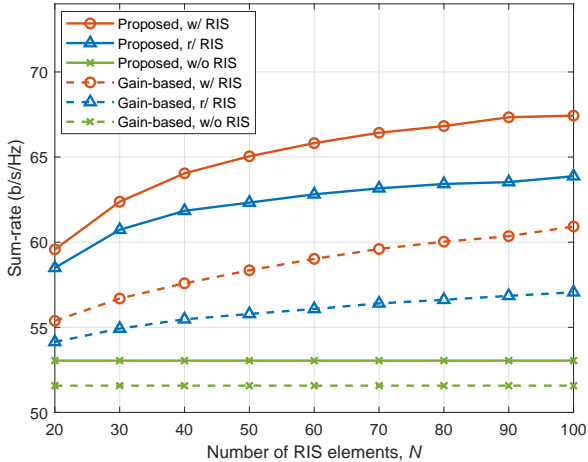}
    \caption{Sum-rate versus the number of RIS reflecting elements, $N$ ($P_{\text{max}} = 20 \text{dBm}$, $M = 32$, $K = 15$).}
    \label{fig:changeN}
\end{figure}

%\begin{figure}[!t]
%	\centering
%	\includegraphics[width = 9 cm]{changeN.eps}
%	\caption{Sum-rate versus the number of RIS reflecting elements, $N$.}
%	\label{fig:changeN}
%\end{figure}
Then, the sum-rate versus the number of RIS reflecting elements $N$ is plotted in Fig. \ref{fig:changeN}. Similar conclusions can be drawn that the deployment of RIS will promote the system performance and the proposed algorithm can achieve better performance than its competitors. Moreover, benefiting from larger gain of passive beamforming provided by more reflecting elements, the sum-rate of all schemes increases with the increasing of $N$.

\begin{figure}[!t]
    \centering
    \includegraphics[width = 3.5 in]{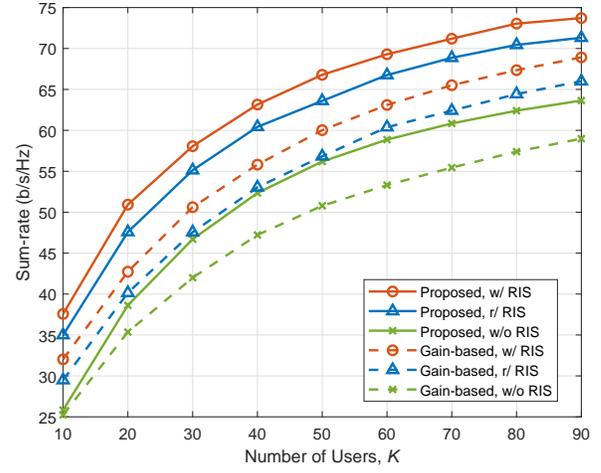}
    \caption{Sum-rate versus the number of users, $K$ ($P_{\text{max}} = 15 \text{dBm}$, $M = 32$, $N = 64$).}
    \label{fig:changeK}
\end{figure}

%\begin{figure}[!t]
%	\centering
%	\includegraphics[width = 9 cm]{changeK.eps}
%	\caption{Sum-rate versus the number of users, $K$.}
%	\label{fig:changeK}
%\end{figure}
Fig. \ref{fig:changeK} describes the sum-rate as a function of the number of users $K$.
We observe that the sum-rate of all schemes increases with increasing $K$, since larger multi-user diversity is brought by more users. In addition, it is obvious that the performance improvement of sum-rate becomes saturated when users are very dense due to limited network resources to allocate to each user and larger multi-user interference.
%Fig. \ref{fig:changeK} describes the sum-rate as a function of the number of users $K$.
%Within a certain range, the higher utilization of resources can be obtained when more users are distributed in the network, thus, the sum-rate of all schemes increases with the increasing of $K$.
%We observe that the sum-rate of all schemes increases with the increasing of $K$, since higher utilization of resources can be obtained when there exists more users within a certain range.
Moreover, our proposed algorithm remains better performance compared with the gain-based algorithm for different numbers of users. %,
%and the performance gap between two algorithms becomes larger with the increasing of $K$.
%Especially when the number of user is $20$, the proposed algorithm under ``w/o RIS" scenario performs close to the gain-based algorithm under ``r/ RIS" scenario.
Since the gain-based user association algorithm only determines the BS-user association based on the direct channel gain, more users will be associated with the closer BSs, i.e., BS2 and BS3. Thus, less power will be allocated to each user associated with these two BSs when more users are distributed in the network, which significantly degrades the sum-rate performance. %causes the worse performance compared with the proposed algorithm. %This fact illustrates the advantage of our proposed joint optimization algorithm.%, which means the proposed algorithm can obtain relatively more sensible BS-user association results in RIS-assisted multi-cell networks., because of the lower path loss
\begin{figure}[!t]
    \centering
    \includegraphics[width = 3.5 in]{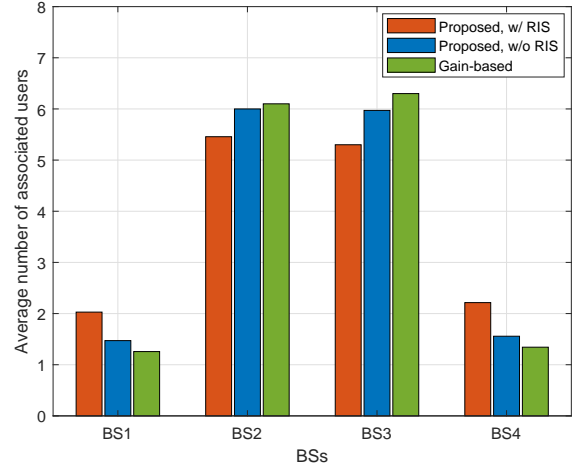}
    \caption{Average number of users served by each BS ($P_{\text{max}} = 20 \text{dBm}$, $M = 32$, $N = 64$, $K = 15$).}
    \label{fig:assonumber}
\end{figure}
Thus, in Fig. \ref{fig:assonumber}, we present the average number of users served by each BS of different algorithms under different scenarios, in which
BS1 and BS4 are at relatively farther distances away from users.
We only plot the ``gain-based" result in this figure since the resulting BS-user association of gain-based algorithm will not be affected by the deployment of RIS, as we analysed before.
It can be easily observed that with the deployment of RIS, the BSs far away from users, i.e., BS1 and BS4, can serve more users than the system without RIS by comparing the ``proposed, w/ RIS" and ``proposed, w/o RIS" schemes. This phenomenon verifies that the deployment of RIS can expand cell coverage in cellular networks.
Moreover, it is obvious that more users will be associated with farther BSs by utilizing the proposed algorithm compared with gain-based algorithm. % since farther BSs will have higher path loss to users.
These facts demonstrate that the deployment of RIS and the proposed joint BS-RIS-user association design can facilitate the load balancing between BSs, which is beneficial to the utilization of resources.%it is obvious that the proposed algorithm will associate more users with the two BSs far from users comparing with the gain-based algorithm, which demonstrates the advantage on load balancing between BSs of the proposed algorithm.

%\begin{figure}[!t]
%	\centering
%	\includegraphics[width = 9cm]{changeBS_assonumber.eps}
%	\caption{Number of associated users versus the distance between BS1/BS4 and the RIS.}
%	\label{fig:bs2}
%\end{figure}

\begin{figure}[!t]
    \centering
    \includegraphics[width = 3.5 in]{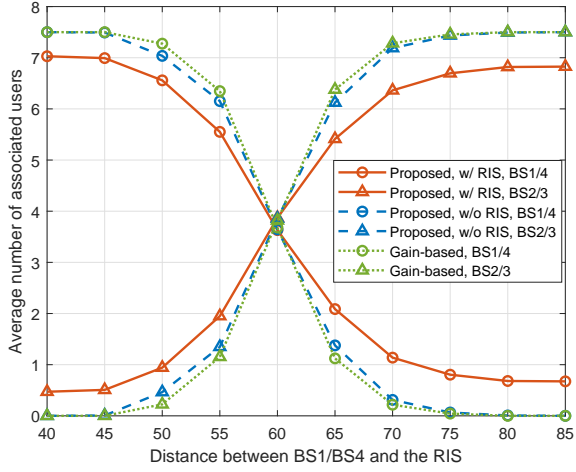}
    \caption{Average number of associated users versus the distance between BS1/BS4 and the RIS ($P_{\text{max}} = 20 \text{dBm}$, $M = 32$, $N = 64$, $K = 15$).}
    \label{fig:bs2}
\end{figure}
To further demonstrate the advantages of our proposed algorithm for load balancing, the average number of associated users versus the distance between BS1/BS4 and RIS is plotted in Fig. \ref{fig:bs2}.
%For simplicity, we assume that BS1 and BS4 are at the same distance from RIS, while BS2 and BS3 also are at the same distance from RIS.
According to the geometry positions presented in Fig. \ref{fig:location}, BS1 and BS4 will achieve the same performance under the same settings, and so will BS2 and BS3.
Thus, we only plot two curves under each scheme in Fig. \ref{fig:bs2}. To be specific, the curve ``BS1/4" is the average number of users associated with BS1 and BS4, while the curve ``BS2/3" is the average number of users associated with BS2 and BS3.
We can observe that the proposed algorithm will associate more users to farther BSs and less users to closer BSs compared with gain-based algorithm in all distance ranges.
%Moreover, the distance between BS2/BS3 ad RIS is set as $60$m, which means all BSs have the same distance to RIS when the distance between BS1/BS4 and RIS is set as.
Moreover, the average number of associated users with different BSs gets closer when the distance from the BSs to the RIS tends to be equal to $60$m.
These facts further verify the load balancing ability of the proposed algorithm and the cell edge expansion by deploying RIS in cellular networks.

%\begin{figure}[!t]
%	\centering
%	\includegraphics[width = 9cm]{changeBS_user.eps}
%	\caption{Sum-rate versus the distance between BS1/BS4 and the RIS.}
%	\label{fig:bs1}
%\end{figure}

%Fig. \ref{fig:bs1} shows the sum-rate versus the distance between BS1/BS4 and RIS. It can be observed that although the proposed algorithm balances the load between BSs, it also achieves better performance compared with the ``Gain-based" algorithm.

\section{Conclusion}
In this paper, we considered a practical RIS-assisted multi-cell wireless network and investigated to maximize the sum-rate by jointly optimizing the BS-RIS-user association, the active beamforming at BSs, as well as the passive beamforming of RIS.
An efficient joint association and beamforming optimization algorithm exploiting FP, MM and ADMM methods was developed to solve this complicated non-convex problem.
Simulation results demonstrated the effectiveness and the performance improvement of our proposed algorithm.
To be specific, the deployment of RIS with the proposed joint association and beamforming design algorithm can expand the cell coverage, facilitate the load balancing between BSs, and consequently promotes the performance of cellular network.
For the future studies, it would be significative to extend the proposed joint BS-RIS-user association and beamforming design architecture to multi-RIS assisted multi-cell multi-user systems. Moreover, the fairness among users is also a worth studying issue as the extended research directions.%the proposed joint association and beamforming design algorithm can facilitate the load balancing between BSs, which is beneficial to the utilization of resources.
%Then, the performance gap between the proposed algorithm and the comparison algorithm also verified that the joint BS-RIS-user association in RIS-assisted cellular network will promote the system performance.
%Moreover, the joint BS-RIS-user association problem is a new but essential in RIS-assisted cellular network, for which more research efforts are needed.
\appendices{
\section{Proof of Proposition 1}
\begin{figure*}[!t]
% ensure that we have normalsize text
\normalsize
% Store the current equation number.
\setcounter{mytempeqncnt}{\value{equation}}
% Set the equation number to one less than the one desired for the first equation here. The value here will have to changed if equations are added or removed prior to the place these equations are referenced in the main text.
\setcounter{equation}{53}
\begin{subequations}
\label{ap:g1h}
\begin{align}
-\mathcal{H}_{\mathbf{w}_{j, k}, \mathbf{w}_{j, k}^{*}}g_{\delta}\left(\mathbf{w}_{j, k}, \mathbf{w}_{j, k}^{*}\right) &= \exp \Big(-\frac{\|\mathbf{w}_{j, k}\|_{2}^{2}}{\delta}\Big)  \Big( \frac{1}{\delta^2}\mathbf{w}_{j,k}\mathbf{w}_{j,k}^H - \frac{1}{\delta}\mathbf{I}_M \Big),\\%= &-\mathcal{D}_{\mathbf{w}_{j, k}}(\mathcal{D}_{\mathbf{w}_{j, k}^{*}}h_{\delta})^T
-\mathcal{H}_{\mathbf{w}_{j, k}^{*}, \mathbf{w}_{j, k}^{*}} g_{\delta}\left(\mathbf{w}_{j, k}, \mathbf{w}_{j, k}^{*}\right)& = \exp \Big(-\frac{\|\mathbf{w}_{j, k}\|_{2}^{2}}{\delta}\Big)  \Big( \frac{1}{\delta^2}\mathbf{w}_{j,k}\mathbf{w}_{j,k}^T \Big),\\% = &-\mathcal{D}_{\mathbf{w}_{j, k}^{*}}(\mathcal{D}_{\mathbf{w}_{j, k}^{*}}h_{\delta})^T
-\mathcal{H}_{\mathbf{w}_{j, k}, \mathbf{w}_{j, k}} g_{\delta}\left(\mathbf{w}_{j, k}, \mathbf{w}_{j, k}^{*}\right) &= \exp \Big(-\frac{\|\mathbf{w}_{j, k}\|_{2}^{2}}{\delta}\Big)  \Big( \frac{1}{\delta^2}\mathbf{w}_{j,k}^{*}\mathbf{w}_{j,k}^H \Big),\\% = &-\mathcal{D}_{\mathbf{w}_{j, k}}(\mathcal{D}_{\mathbf{w}_{j, k}}h_{\delta})^T
-\mathcal{H}_{\mathbf{w}_{j, k}^{*}, \mathbf{w}_{j, k}} g_{\delta}\left(\mathbf{w}_{j, k}, \mathbf{w}_{j, k}^{*}\right) &= \exp \Big(-\frac{\|\mathbf{w}_{j, k}\|_{2}^{2}}{\delta}\Big)  \Big( \frac{1}{\delta^2}\mathbf{w}_{j,k}^{*}\mathbf{w}_{j,k}^T - \frac{1}{\delta}\mathbf{I}_M \Big).%= &-\mathcal{D}_{\mathbf{w}_{j, k}^{*}}(\mathcal{D}_{\mathbf{w}_{j, k}}h_{\delta})^T
\end{align}
\end{subequations}
% Restore the current equation number.
\setcounter{equation}{\value{mytempeqncnt}}
% IEEE uses as a separator
\hrulefill
% The spacer can be tweaked to stop underfull vboxes.
\vspace*{4pt}
\end{figure*}
\begin{figure*}
% ensure that we have normalsize text
\normalsize
% Store the current equation number.
\setcounter{mytempeqncnt}{\value{equation}}
% Set the equation number to one less than the one desired for the first equation here. The value here will have to changed if equations are added or removed prior to the place these equations are referenced in the main text.
\setcounter{equation}{61}
\begin{subequations}
\label{ap:g2h}
%\begin{small}
\begin{align}
-\mathcal{H}_{\boldsymbol{\varphi}_j, \boldsymbol{\varphi}_j^{*}}g_{\delta}(\boldsymbol{\varphi}_j, \boldsymbol{\varphi}_j^{*}) = & \varepsilon_j \Big( \frac{1}{\delta^2}\left((\ln(\boldsymbol{\varphi}_j)) \oslash \boldsymbol{\varphi}_j^*\right)\left((\ln(\boldsymbol{\varphi}_j))^H \oslash \boldsymbol{\varphi}_j^T\right) - \frac{1}{\delta}\operatorname{diag}\{\mathbf{1}_N \oslash \boldsymbol{\varphi}_j \oslash \boldsymbol{\varphi}_j^{*} \} \Big),\\%\mathcal{D}_{\boldsymbol{\varphi}_j}\left(\mathcal{D}_{\boldsymbol{\varphi}_j^{*}}h_2\right)^T =\exp(-\frac{\big\|\ln(\boldsymbol{\varphi}_{j})\big\|_2^2}{\delta})
-\mathcal{H}_{\boldsymbol{\varphi}_j^{*}, \boldsymbol{\varphi}_j^{*}} g_{\delta}(\boldsymbol{\varphi}_j, \boldsymbol{\varphi}_j^{*}) = & \varepsilon_j \Big( \frac{1}{\delta^2}\left((\ln(\boldsymbol{\varphi}_j)) \oslash \boldsymbol{\varphi}_j^*\right)\left((\ln(\boldsymbol{\varphi}_j))^T \oslash \boldsymbol{\varphi}_j^H\right)- \frac{1}{\delta}\operatorname{diag}\{\ln(\boldsymbol{\varphi}_{j}) \oslash \boldsymbol{\varphi}_j^* \oslash \boldsymbol{\varphi}_j^{*} \} \Big),\\%\mathcal{D}_{\boldsymbol{\varphi}_j^{*}}\left(\mathcal{D}_{\boldsymbol{\varphi}_j^{*}}h_2\right)^T =\exp(-\frac{\big\|\ln(\boldsymbol{\varphi}_{j})\big\|_2^2}{\delta})
-\mathcal{H}_{\boldsymbol{\varphi}_j, \boldsymbol{\varphi}_j} g_{\delta}(\boldsymbol{\varphi}_j, \boldsymbol{\varphi}_j^{*}) = & \varepsilon_j \Big( \frac{1}{\delta^2}\left((\ln(\boldsymbol{\varphi}_j^*)) \oslash \boldsymbol{\varphi}_j\right)\left((\ln(\boldsymbol{\varphi}_j))^H \oslash \boldsymbol{\varphi}_j^T\right) - \frac{1}{\delta}\operatorname{diag}\{\ln(\boldsymbol{\varphi}_{j}^*) \oslash \boldsymbol{\varphi}_j \oslash \boldsymbol{\varphi}_j \} \Big),\\%\mathcal{D}_{\boldsymbol{\varphi}_j}\left(\mathcal{D}_{\boldsymbol{\varphi}_j}h_2\right)^T =\exp(-\frac{\big\|\ln(\boldsymbol{\varphi}_{j})\big\|_2^2}{\delta})
-\mathcal{H}_{\boldsymbol{\varphi}_j^{*}, \boldsymbol{\varphi}_j} g_{\delta}(\boldsymbol{\varphi}_j, \boldsymbol{\varphi}_j^{*}) = &\varepsilon_j \Big( \frac{1}{\delta^2}\left((\ln(\boldsymbol{\varphi}_j^*)) \oslash \boldsymbol{\varphi}_j\right)\left((\ln(\boldsymbol{\varphi}_j))^T \oslash \boldsymbol{\varphi}_j^H\right) - \frac{1}{\delta}\operatorname{diag}\{\mathbf{1}_N \oslash \boldsymbol{\varphi}_j^* \oslash \boldsymbol{\varphi}_j \} \Big).%\mathcal{D}_{\boldsymbol{\varphi}_j^{*}}\left(\mathcal{D}_{\boldsymbol{\varphi}_j}h_2\right)^T =\exp(-\frac{\big\|\ln(\boldsymbol{\varphi}_{j})\big\|_2^2}{\delta})
\end{align}
%\end{small}
\end{subequations}
% Restore the current equation number.
\setcounter{equation}{\value{mytempeqncnt}}
% IEEE uses as a separator
\hrulefill
% The spacer can be tweaked to stop underfull vboxes.
\vspace*{4pt}
\end{figure*}
%For inequality (a) in \eqref{eq:w_relax}, under the MM algorithm framework, we aim to find a convex upper-bound $g_1\left(\mathbf{w}_{j, k}\right)$ which satisfies
%\begin{equation}
%-h_1 \left(\mathbf{w}_{j, k}\right) \le g_1\left(\mathbf{w}_{j, k}\right),
%~ \forall j \in \mathcal{J},   ~ \forall k \in \mathcal{K},
%\end{equation}
%to transform the inequality to a more tractable constraint.
Since $-g_{\delta}(\|\mathbf{w}_{j, k}\|_2^2)$ is twice differentiable, inspired by \cite{Y. Sun}, the second-order Taylor expansion is utilized to construct its upper-bound $\widetilde{g}_{\delta}(\mathbf{w}_{j, k}|\mathbf{w}_{j, k}^{(t-1)})$. As introduced in \cite{A. Hjorungnes}, the second-order Taylor series of function $-g_{\delta}\left(\|\mathbf{w}_{j, k}\|_2^2\right)$ at point $\mathbf{w}_{j, k}$ can be written as
\begin{equation}
\label{ap:hh1}
\begin{aligned}
H_1\left(\mathbf{w}_{j, k}\right)  \hspace{-0.05cm}= \hspace{-0.05cm} &-g_{\delta}(\|\mathbf{w}_{j, k}^{(t-1)}\|_2^2)  \hspace{-0.05cm}- \hspace{-0.05cm} \left(\mathcal{D}_{\mathbf{w}_{j, k}} g_{\delta}(\mathbf{w}_{j, k}, \mathbf{w}_{j, k}^{*})\right) d \mathbf{w}_{j, k}\\
&-\left(\mathcal{D}_{\mathbf{w}_{j, k}^{*}} g_{\delta} (\mathbf{w}_{j, k}, \mathbf{w}_{j, k}^{*})\right) d \mathbf{w}_{j, k}^{*}\\
&+\frac{1}{2}\left[d \mathbf{w}_{j, k}^{H} ~~ d \mathbf{w}_{j, k}^T\right] \mathcal{H}_{j,k} \left[\begin{array}{c}
d \mathbf{w}_{j, k} \\
d \mathbf{w}_{j, k}^{*}
\end{array}\right],
\end{aligned}
\end{equation}
in which $\mathcal{D}_{\mathbf{w}_{j, k}} g_{\delta}(\mathbf{w}_{j, k}, \mathbf{w}_{j, k}^{*})$ and $\mathcal{D}_{\mathbf{w}_{j, k}^{*}} g_{\delta}(\mathbf{w}_{j, k}, \mathbf{w}_{j, k}^{*})$ represent the complex-valued derivatives of the function $g_{\delta}(\mathbf{w}_{j, k}, \mathbf{w}_{j, k}^{*})$ with respect to $\mathbf{w}_{j, k}$ and $\mathbf{w}_{j, k}^{*}$, respectively. As introduced in Lemma 5.2 \cite{A. Hjorungnes}, the complex Hessian matrix of $-g_{\delta} \left(\mathbf{w}_{j, k}\right)$ can be expressed as
\begin{equation}
\mathcal{H}_{j,k} \hspace{-0.1cm}=  \hspace{-0.1cm}\left[\begin{array}{cc}
-\mathcal{H}_{\mathbf{w}_{j, k}, \mathbf{w}_{j, k}^{*}}g_{\delta}  & -\mathcal{H}_{\mathbf{w}_{j, k}^{*}, \mathbf{w}_{j, k}^{*}} g_{\delta} \\
-\mathcal{H}_{\mathbf{w}_{j, k}, \mathbf{w}_{j, k}} g_{\delta} & -\mathcal{H}_{\mathbf{w}_{j, k}^{*}, \mathbf{w}_{j, k}} g_{\delta}
\end{array}\right],
\end{equation}
where $\mathcal{H}_{\mathbf{w}_{j, k}, \mathbf{w}_{j, k}^{*}}g_{\delta}\triangleq \mathcal{D}_{\mathbf{w}_{j, k}}(\mathcal{D}_{\mathbf{w}_{j, k}^{*}}g_{\delta}(\mathbf{w}_{j, k}, \mathbf{w}_{j, k}^{*}))^T$ represents the complex Hessian matrix for function $g_{\delta}(\mathbf{w}_{j, k}, \mathbf{w}_{j, k}^{*})$ with respect to $(\mathbf{w}_{j, k},\mathbf{w}_{j, k}^{*})$, and similar definitions for the other matrices in $\mathcal{H}_{j,k}$.
Based on the definition
\begin{equation}
-g_{\delta}\left(\|\mathbf{w}_{j, k}\|_2^2\right) = \exp\big(-\frac{\left\|\mathbf{w}_{j, k}\right\|_2^2}{\delta}\big)-1,
\end{equation}
\vspace{-0.2cm}
\hspace{-0.12cm}we can obtain the derivatives
\begin{subequations}
\label{eq:g1d}
\begin{align}
-\mathcal{D}_{\mathbf{w}_{j, k}} g_{\delta}\left(\mathbf{w}_{j, k}, \mathbf{w}_{j, k}^{*}\right)\hspace{-0.1cm}=\hspace{-0.1cm}&-\hspace{-0.1cm}\frac{1}{\delta}\exp \Big(-\frac{\left\|\mathbf{w}_{j, k}\right\|_{2}^{2}}{\delta}\Big)  \mathbf{w}_{j , k}^{H},\\
-\mathcal{D}_{\mathbf{w}_{j, k}^{*}} g_{\delta}\left(\mathbf{w}_{j, k}, \mathbf{w}_{j, k}^{*}\right)\hspace{-0.1cm}=\hspace{-0.1cm}&-\hspace{-0.1cm}\frac{1}{\delta}\exp \Big(-\frac{\left\|\mathbf{w}_{j, k}\right\|_{2}^{2}}{\delta}\Big)  \mathbf{w}_{j , k}^{T},
\end{align}
\end{subequations}
and Hessian matrices \eqref{ap:g1h} as shown on the top of next page.

%$respectively.

Then, to construct an upper-bound of $H_1\left(\mathbf{w}_{j, k}\right)$ at point $\mathbf{w}_{j, k}$, a matrix $\widetilde{\mathbf{M}}_{j,k}$, which satisfies $\widetilde{\mathbf{M}}_{j,k} \succeq  \mathcal{H}_{j,k}$, needs to be adopted.
%\begin{equation}
%\label{ap:g1}
%\begin{aligned}
%g_1\left(\mathbf{w}_{j, k}\right) = &-h_1 \left(\mathbf{w}_{j, k}^{(t-1)}\right) - \left(\mathcal{D}_{\mathbf{w}_{j, k}} h_1\left(\mathbf{w}_{j, k}, \mathbf{w}_{j, k}^{*}\right)\right) d \mathbf{w}_{j, k}\\
%&-\left(\mathcal{D}_{\mathbf{w}_{j, k}^{*}} h_1 \left(\mathbf{w}_{j, k}, \mathbf{w}_{j, k}^{*}\right)\right) d \mathbf{w}_{j, k}^{*}\\
%&+\frac{1}{2}\left[d \mathbf{w}_{j, k}^{H} ~~ d \mathbf{w}_{j, k}^T\right] \tilde{\mathbf{M}}_{j,k} \left[\begin{array}{c}
%d \mathbf{w}_{j, k} \\
%d \mathbf{w}_{j, k}^{*}
%\end{array}\right],
%\end{aligned}
%\end{equation}
Thus, we set
\setcounter{equation}{54}
\begin{equation}
\widetilde{\mathbf{M}}_{j,k} = \left[\begin{array}{cc}
\mathbf{M}_{j,k}  & \mathbf{0}_{M \times M} \\
\mathbf{0}_{M \times M} & \mathbf{M}_{j,k}^T
\end{array}\right] = \max{\{0,\lambda_{j,k}\}} \cdot \mathbf{I}_{2M },
\end{equation}
where $\lambda_{j,k}$ is the maximum eigenvalue of the matrix $\mathcal{H}_{j,k} $.
Moreover, $d \mathbf{w}_{j, k}$ equals to $\mathbf{w}_{j, k} - \mathbf{w}_{j, k}^{(t-1)}$.
Finally, %by substituting the equations \eqref{eq:g1d} and \eqref{ap:g1h} into \eqref{ap:g1},
a convex upper-bound $\widetilde{g}_{\delta}(\mathbf{w}_{j, k}|\mathbf{w}_{j, k}^{(t-1)})$ can be written as
\begin{equation}
\begin{aligned}
\widetilde{g}_{\delta}(\mathbf{w}_{j, k}|\mathbf{w}_{j, k}^{(t-1)}) = &\mathfrak{Re} \{\mathbf{w}_{j,k}^H \mathbf{M}_{j,k} \mathbf{w}_{j,k} \} + \mathfrak{Re} \{ \boldsymbol{\iota}_{j,k}^H \mathbf{w}_{j,k}\}\\
&+ \varpi_{j,k},  ~ \forall j ,  ~ \forall k ,
\end{aligned}
\end{equation}
in which
\begin{subequations}
%\begin{small}
\begin{align}
\mathbf{M}_{j,k}  \triangleq &\max{\{0,\lambda_{j,k}\}} \cdot \mathbf{I}_M,  ~ \forall j,  \forall k,\\
\beta_{j,k}^{(t-1)} \triangleq & \frac{1}{\delta}\exp \Big(-\frac{\big\|\mathbf{w}_{j, k}^{(t-1)}\big\|_{2}^{2}}{\delta}\Big),  ~ \forall j,  \forall k ,\\
\boldsymbol{\iota}_{j,k}^H  \triangleq&-\hspace{-0.1cm}2(\mathbf{w}_{j,k}^{(t-1)})^H\mathbf{M}_{j,k} \hspace{-0.1cm}- \hspace{-0.1cm} 2\beta_{j,k}^{(t-1)}(\mathbf{w}_{j,k}^{(t-1)})^H,~  \forall j , \forall k ,\\
\varpi_{j,k}  \triangleq&- h_{\delta}(\|\mathbf{w}_{j, k}^{(t-1)}\|_2^2)+\mathfrak{Re} \{ (\mathbf{w}_{j,k}^{(t-1)})^H \mathbf{M}_{j,k} \mathbf{w}_{j,k}^{(t-1)}\}  \hspace{-0.05cm}\\ \nonumber
&+ \hspace{-0.05cm}2\mathfrak{Re} \{ \beta_{j,k}^{(t-1)} (\mathbf{w}_{j,k}^{(t-1)})^H\mathbf{w}_{j,k}^{(t-1)}\}
, \forall j ,  \forall k .
\end{align}
%\end{small}
\end{subequations}

Hence, Proposition 1 is proved.\hfill $\blacksquare$
\section{Proof of Proposition 2}
Since $-g_{\delta}\left(\|\ln(\boldsymbol{\varphi}_{j})\|_2^2\right)$ is twice differentiable, the second-order Taylor expansion of $-g_{\delta}\left(\|\ln(\boldsymbol{\varphi}_{j})\|_2^2\right)$ at point $\boldsymbol{\varphi}_j$ can be adopted as
\begin{equation}
\label{eq:hh2}
\begin{aligned}
H_2(\boldsymbol{\varphi}_j) = &-g_{\delta}\left(\|\ln(\boldsymbol{\varphi}_{j})^{(t-1)}\|_2^2\right) -(\mathcal{D}_{\boldsymbol{\varphi}_j} g_{\delta}(\boldsymbol{\varphi}_j, \boldsymbol{\varphi}_j^{*})) d \boldsymbol{\varphi}_j\\
&-(\mathcal{D}_{\boldsymbol{\varphi}_j^{*}} g_{\delta} (\boldsymbol{\varphi}_j, \boldsymbol{\varphi}_j^{*})) d \boldsymbol{\varphi}_j^{*}\\
&+\frac{1}{2}\left[d \boldsymbol{\varphi}_j^{H} ~~ d \boldsymbol{\varphi}_j^T\right] \mathcal{H}_{j} \left[\begin{array}{c}
d \boldsymbol{\varphi}_j \\
d \boldsymbol{\varphi}_j^{*}
\end{array}\right],
\end{aligned}
\end{equation}
in which similar definitions for the derivatives and Hessian matrices as that in \eqref{ap:hh1} are used, and
\begin{equation}
\mathcal{H}_{j} = \left[\begin{array}{cc}
-\mathcal{H}_{\boldsymbol{\varphi}_j, \boldsymbol{\varphi}_j^{*}}g_{\delta}  & -\mathcal{H}_{\boldsymbol{\varphi}_j^{*}, \boldsymbol{\varphi}_j^{*}} g_{\delta} \\
-\mathcal{H}_{\boldsymbol{\varphi}_j, \boldsymbol{\varphi}_j} g_{\delta} & -\mathcal{H}_{\boldsymbol{\varphi}_j^{*}, \boldsymbol{\varphi}_j} g_{\delta}
\end{array}\right],
\end{equation}
is the complex Hessian matrix of $-g_{\delta}(\|\ln(\boldsymbol{\varphi}_{j})\|_2^2)$.
Then, based on the defination
\begin{equation}
-g_{\delta}(\|\ln(\boldsymbol{\varphi}_{j})\|_2^2) = \exp\Big(-\frac{\|\ln(\boldsymbol{\varphi}_{j})\|_2^2}{\delta}\Big) - 1,
\end{equation}
we can obtain
\begin{subequations}
\label{eq:g2d}
\begin{align}
-\mathcal{D}_{\boldsymbol{\varphi}_j} g_{\delta}(\boldsymbol{\varphi}_j, \boldsymbol{\varphi}_j^{*})\hspace{-0.05cm}=\hspace{-0.05cm}&-\hspace{-0.05cm}\frac{1}{\delta} \hspace{-0.05cm} \exp(-\frac{\|\ln(\boldsymbol{\varphi}_{j})\|_2^2}{\delta})  ((\ln(\boldsymbol{\varphi}_j))^H \hspace{-0.05cm}\oslash\hspace{-0.05cm} \boldsymbol{\varphi}_j^T),\\
\hspace{-0.2cm}-\mathcal{D}_{\boldsymbol{\varphi}_j^{*}} g_{\delta}(\boldsymbol{\varphi}_j, \boldsymbol{\varphi}_j^{*})\hspace{-0.05cm}=\hspace{-0.05cm}&-\hspace{-0.05cm}\frac{1}{\delta}\hspace{-0.05cm}\exp(-\frac{\|\ln(\boldsymbol{\varphi}_{j})\|_2^2}{\delta})   ((\ln(\boldsymbol{\varphi}_j))^T \hspace{-0.05cm}\oslash\hspace{-0.05cm} \boldsymbol{\varphi}_j^H),
\end{align}
\end{subequations}and \eqref{ap:g2h} as shown on the top of this page,
in which $\oslash$ presents the element-wise division and for brevity we define
\setcounter{equation}{62}
\begin{equation}
\varepsilon_j \triangleq \frac{1}{\delta}\exp\Big(-\frac{\big\|\ln(\boldsymbol{\varphi}_{j})\big\|_2^2}{\delta}\Big).
\end{equation}
%\begin{equation}
%\label{ap:g2}
%\begin{aligned}
%g_2(\boldsymbol{\varphi}_j) = &-h_2 \big(\boldsymbol{\varphi}_j^{(t-1)}\big) + -(\mathcal{D}_{\boldsymbol{\varphi}_j} h_2(\boldsymbol{\varphi}_j, \boldsymbol{\varphi}_j^{*})) d \boldsymbol{\varphi}_j\\
%&-(\mathcal{D}_{\boldsymbol{\varphi}_j^{*}} h_2 (\boldsymbol{\varphi}_j, \boldsymbol{\varphi}_j^{*})) d \boldsymbol{\varphi}_j^{*}\\
%&+\frac{1}{2}\left[d \boldsymbol{\varphi}_j^{H} ~~ d \boldsymbol{\varphi}_j^T\right] \tilde{\boldsymbol{\Xi}}_{j} \left[\begin{array}{c}
%d \boldsymbol{\varphi}_j \\
%d \boldsymbol{\varphi}_j^{*}
%\end{array}\right],
%\end{aligned}
%\end{equation}
To construct the upper-bound $\overline{g}_{\delta}(\boldsymbol{\varphi}_j|\boldsymbol{\varphi}_j^{(t-1)})$, the matrix
\begin{equation}
\widetilde{\boldsymbol{\Xi}}_{j} = \left[\begin{array}{cc}
\boldsymbol{\Xi}_{j}  & \mathbf{0}_{N \times N} \\
\mathbf{0}_{N \times N} & \boldsymbol{\Xi}_{j}^T
\end{array}\right] = \max{\{0,\lambda_{j}\}}\cdot\mathbf{I}_{2N },
\end{equation}
is utilized as the upper-bound of the Hessian matrix $\mathcal{H}_{j}$ in \eqref{eq:hh2}, which satisfies the constraint $\widetilde{\boldsymbol{\Xi}}_{j} \succeq  \mathcal{H}_{j}$.
Here, $\lambda_{j}$ is the maximum eigenvalue of the matrix $\mathcal{H}_{j} $.
Finally, by substituting \eqref{eq:g2d}, \eqref{ap:g2h}, and $d \boldsymbol{\varphi}_j = \boldsymbol{\varphi}_j - \boldsymbol{\varphi}_j^{(t-1)}$ into \eqref{eq:hh2}, the upper-bound $\overline{g}_{\delta}(\boldsymbol{\varphi}_j|\boldsymbol{\varphi}_j^{(t-1)})$ can be given by
\begin{equation}
\label{eq:g2f}
\begin{aligned}
\hspace{-0.3cm}\overline{g}_{\delta}(\boldsymbol{\varphi}_j|\boldsymbol{\varphi}_j^{(t-1)}) = \mathfrak{Re} \{\boldsymbol{\varphi}_j^H \boldsymbol{\Xi}_{j} \boldsymbol{\varphi}_j \} \hspace{-0.05cm}+ \hspace{-0.05cm}\mathfrak{Re} \{ \boldsymbol{\kappa}_{j}^H\boldsymbol{\varphi}_j\}\hspace{-0.05cm} +\hspace{-0.05cm} \mu_{j},\forall j,
\end{aligned}
\end{equation}
in which
\begin{subequations}
%\begin{small}
\begin{align}
\hspace{-1cm}\mathbf{\Xi}_j \triangleq& \max{\{0,\lambda_{j}\}} \cdot \mathbf{I}_N, ~ \forall j ,\\
\boldsymbol{\kappa}_{j} \triangleq&-2(\boldsymbol{\varphi}_j^{(t-1)})^H\boldsymbol{\Xi}_{j} - 2\varepsilon_j^{(t-1)} (\boldsymbol{\phi}_j^{(t-1)})^H ,  ~ \forall j ,  \\
\mu_{j} \triangleq&- g_{\delta}(\|\ln(\boldsymbol{\varphi}_{j}^{(t-1)})\|_2^2)\hspace{-0.05cm}+\hspace{-0.05cm}\mathfrak{Re} \{ (\boldsymbol{\varphi}_{j}^{(t-1)})^H \boldsymbol{\Xi}_{j} \boldsymbol{\varphi}_{j}^{(t-1)}\}\hspace{-0.05cm} \\ \nonumber
&+ \hspace{-0.05cm} 2\mathfrak{Re} \{\varepsilon_j^{(t-1)} (\boldsymbol{\phi}_j^{(t-1)})^H \boldsymbol{\varphi}_{j}^{(t-1)}  \}
, ~ \forall j .
\end{align}
%\end{small}
\end{subequations}\hspace{-0.15cm}Here, for brevity we define
\begin{subequations}
\begin{align}
\varepsilon_j^{(t-1)} \triangleq& \frac{1}{\delta}\exp\Big(-\frac{\big\|\ln(\boldsymbol{\varphi}_{j}^{(t-1)})\big\|_2^2}{\delta}\Big), ~ \forall j,\\
(\boldsymbol{\phi}_j^{(t-1)})^H \triangleq&\ln({\boldsymbol{\varphi}_{j}^{(t-1)})}^H \oslash {(\boldsymbol{\varphi}_{j}^{(t-1)}) }^T, ~ \forall j .
\end{align}
\end{subequations}

Now, the proof of Proposition 2 is complete.\hfill $\blacksquare$
\section{Proof of Proposition 3}
\begin{figure*}
% ensure that we have normalsize text
\normalsize
% Store the current equation number.
\setcounter{mytempeqncnt}{\value{equation}}
% Set the equation number to one less than the one desired for the first equation here. The value here will have to changed if equations are added or removed prior to the place these equations are referenced in the main text.
\setcounter{equation}{70}
\begin{subequations}
\label{ap:g3h}
%\begin{small}
\begin{align}
\mathcal{H}_{\boldsymbol{\varphi}_j, \boldsymbol{\varphi}_j^{*}}g_{\delta}(\boldsymbol{\varphi}_j, \boldsymbol{\varphi}_j^{*}) = & \varepsilon_j \Big( \frac{1}{\delta}\operatorname{diag}\{\mathbf{1}_N \oslash \boldsymbol{\varphi}_j \oslash \boldsymbol{\varphi}_j^{*} \} -
\frac{1}{\delta^2}\left((\ln(\boldsymbol{\varphi}_j)) \oslash \boldsymbol{\varphi}_j^*\right)\left((\ln(\boldsymbol{\varphi}_j))^H \oslash \boldsymbol{\varphi}_j^T\right)\Big),\\%\mathcal{D}_{\boldsymbol{\varphi}_j}\left(\mathcal{D}_{\boldsymbol{\varphi}_j^{*}}h_2\right)^T =\exp(-\frac{\big\|\ln(\boldsymbol{\varphi}_{j})\big\|_2^2}{\delta})
\mathcal{H}_{\boldsymbol{\varphi}_j^{*}, \boldsymbol{\varphi}_j^{*}} g_{\delta}(\boldsymbol{\varphi}_j, \boldsymbol{\varphi}_j^{*}) = & \varepsilon_j \Big(\frac{1}{\delta}\operatorname{diag}\{\ln(\boldsymbol{\varphi}_{j}) \oslash \boldsymbol{\varphi}_j^* \oslash \boldsymbol{\varphi}_j^{*} \}
-\frac{1}{\delta^2}\left((\ln(\boldsymbol{\varphi}_j)) \oslash \boldsymbol{\varphi}_j^*\right)\left((\ln(\boldsymbol{\varphi}_j))^T \oslash \boldsymbol{\varphi}_j^H\right)\Big),\\%\mathcal{D}_{\boldsymbol{\varphi}_j^{*}}\left(\mathcal{D}_{\boldsymbol{\varphi}_j^{*}}h_2\right)^T =\exp(-\frac{\big\|\ln(\boldsymbol{\varphi}_{j})\big\|_2^2}{\delta})
\mathcal{H}_{\boldsymbol{\varphi}_j, \boldsymbol{\varphi}_j} g_{\delta}(\boldsymbol{\varphi}_j, \boldsymbol{\varphi}_j^{*}) = &\varepsilon_j \Big( \frac{1}{\delta}\operatorname{diag}\{\ln(\boldsymbol{\varphi}_{j}^*) \oslash \boldsymbol{\varphi}_j \oslash \boldsymbol{\varphi}_j \} -\frac{1}{\delta^2}\left((\ln(\boldsymbol{\varphi}_j^*)) \oslash \boldsymbol{\varphi}_j\right)\left((\ln(\boldsymbol{\varphi}_j))^H \oslash \boldsymbol{\varphi}_j^T\right)\Big),\\%\mathcal{D}_{\boldsymbol{\varphi}_j}\left(\mathcal{D}_{\boldsymbol{\varphi}_j}h_2\right)^T =\exp(-\frac{\big\|\ln(\boldsymbol{\varphi}_{j})\big\|_2^2}{\delta})
\mathcal{H}_{\boldsymbol{\varphi}_j^{*}, \boldsymbol{\varphi}_j} g_{\delta}(\boldsymbol{\varphi}_j, \boldsymbol{\varphi}_j^{*}) = & \varepsilon_j \Big( \frac{1}{\delta}\operatorname{diag}\{\mathbf{1}_N \oslash \boldsymbol{\varphi}_j^* \oslash \boldsymbol{\varphi}_j \} -\frac{1}{\delta^2}\left((\ln(\boldsymbol{\varphi}_j^*)) \oslash \boldsymbol{\varphi}_j\right)\left((\ln(\boldsymbol{\varphi}_j))^T \oslash \boldsymbol{\varphi}_j^H\right) \Big).%\mathcal{D}_{\boldsymbol{\varphi}_j^{*}}\left(\mathcal{D}_{\boldsymbol{\varphi}_j}h_2\right)^T =\exp(-\frac{\big\|\ln(\boldsymbol{\varphi}_{j})\big\|_2^2}{\delta})
\end{align}
%\end{small}
\end{subequations}
% Restore the current equation number.
\setcounter{equation}{\value{mytempeqncnt}}
% IEEE uses as a separator
\hrulefill
% The spacer can be tweaked to stop underfull vboxes.
\vspace*{4pt}
\end{figure*}
Similar to the method utilized to construct $\overline{g}_{\delta}(\boldsymbol{\varphi}_j|\boldsymbol{\varphi}_j^{(t-1)})$, we can construct a convex function $\breve{g}_{\delta}(\boldsymbol{\varphi}_j|\boldsymbol{\varphi}_j^{(t-1)})$ as the upper-bound of $g_{\delta}\left(\|\ln(\boldsymbol{\varphi}_{j})\|_2^2\right)$ at point $\boldsymbol{\varphi}_j$ based on the second-order Taylor expansion, which is written as
\begin{equation}
\label{eq:hh3}
\begin{aligned}
H_3(\boldsymbol{\varphi}_j) = &g_{\delta}(\|\ln(\boldsymbol{\varphi}_{j}^{(t-1)})\|_2^2) + \mathcal{D}_{\boldsymbol{\varphi}_j} g_{\delta}(\boldsymbol{\varphi}_j, \boldsymbol{\varphi}_j^{*}) d \boldsymbol{\varphi}_j\\
&+\mathcal{D}_{\boldsymbol{\varphi}_j^{*}} g_{\delta} (\boldsymbol{\varphi}_j, \boldsymbol{\varphi}_j^{*}) d \boldsymbol{\varphi}_j^{*}\\
&+\frac{1}{2}\left[d \boldsymbol{\varphi}_j^{H} ~~ d \boldsymbol{\varphi}_j^T\right] \mathcal{H}_{j} \left[\begin{array}{c}
d \boldsymbol{\varphi}_j \\
d \boldsymbol{\varphi}_j^{*}
\end{array}\right],
\end{aligned}
\end{equation}
in which
\begin{equation}
\mathcal{H}_{j} = \left[\begin{array}{cc}
\mathcal{H}_{\boldsymbol{\varphi}_j, \boldsymbol{\varphi}_j^{*}}g_{\delta}  & \mathcal{H}_{\boldsymbol{\varphi}_j^{*}, \boldsymbol{\varphi}_j^{*}} g_{\delta}\\
\mathcal{H}_{\boldsymbol{\varphi}_j, \boldsymbol{\varphi}_j} g_{\delta} & \mathcal{H}_{\boldsymbol{\varphi}_j^{*}, \boldsymbol{\varphi}_j} g_{\delta}
\end{array}\right],
\end{equation}
is the complex Hessian matrix of $g_{\delta}\left(\|\ln(\boldsymbol{\varphi}_{j})\|_2^2\right)$.
Then, %based on the defination
%\begin{equation}
%h_{\delta}\left(\|\ln(\boldsymbol{\varphi}_{j})\|_2^2\right) = 1- \exp\Big(-\frac{\|\ln(\boldsymbol{\varphi}_{j})\|_2^2}{\delta}\Big),
%\end{equation}
we can obtain the complex-valued derivative and Hessian results of $g_{\delta}\left(\|\ln(\boldsymbol{\varphi}_{j})\|_2^2\right)$ by
\begin{subequations}
\label{eq:g3d}
%\begin{small}
\begin{align}
\hspace{-0.2cm}\mathcal{D}_{\boldsymbol{\varphi}_j} g_{\delta}(\boldsymbol{\varphi}_j, \boldsymbol{\varphi}_j^{*})\hspace{-0.05cm}=\hspace{-0.05cm}&\frac{1}{\delta} \hspace{-0.05cm}\exp(-\frac{\|\ln(\boldsymbol{\varphi}_{j})\|_2^2}{\delta})  ((\ln(\boldsymbol{\varphi}_j))^H \oslash \boldsymbol{\varphi}_j^T),\\
\hspace{-0.2cm}\mathcal{D}_{\boldsymbol{\varphi}_j^{*}} g_{\delta}(\boldsymbol{\varphi}_j, \boldsymbol{\varphi}_j^{*})\hspace{-0.05cm}=\hspace{-0.05cm}&\frac{1}{\delta}\hspace{-0.05cm}\exp(-\frac{\|\ln(\boldsymbol{\varphi}_{j})\|_2^2}{\delta})  ((\ln(\boldsymbol{\varphi}_j))^T \oslash \boldsymbol{\varphi}_j^H),
\end{align}
%\end{small}
\end{subequations}\hspace{-0.15cm}and \eqref{ap:g3h} as shown on the top of this page.
%\begin{equation}
%\label{ap:g2}
%\begin{aligned}
%g_2(\boldsymbol{\varphi}_j) = &-h_2 \big(\boldsymbol{\varphi}_j^{(t-1)}\big) + -(\mathcal{D}_{\boldsymbol{\varphi}_j} h_2(\boldsymbol{\varphi}_j, \boldsymbol{\varphi}_j^{*})) d \boldsymbol{\varphi}_j\\
%&-(\mathcal{D}_{\boldsymbol{\varphi}_j^{*}} h_2 (\boldsymbol{\varphi}_j, \boldsymbol{\varphi}_j^{*})) d \boldsymbol{\varphi}_j^{*}\\
%&+\frac{1}{2}\left[d \boldsymbol{\varphi}_j^{H} ~~ d \boldsymbol{\varphi}_j^T\right] \tilde{\boldsymbol{\Xi}}_{j} \left[\begin{array}{c}
%d \boldsymbol{\varphi}_j \\
%d \boldsymbol{\varphi}_j^{*}
%\end{array}\right],
%\end{aligned}
%\end{equation}
To construct the upper-bound $\breve{g}_{\delta}(\boldsymbol{\varphi}_j|\boldsymbol{\varphi}_j^{(t-1)})$, the matrix
\setcounter{equation}{71}
\begin{equation}
\widetilde{\boldsymbol{\Gamma}}_{j} = \left[\begin{array}{cc}
\boldsymbol{\Gamma}_{j}  & \mathbf{0}_{N \times N} \\
\mathbf{0}_{N \times N} & \boldsymbol{\Gamma}_{j}^T
\end{array}\right] = \max{\{0,\lambda_{j}\}}\cdot\mathbf{I}_{2N },
\end{equation}
is utilized as the upper-bound of Hessian matrix $\mathcal{H}_{j}$ in \eqref{eq:hh3}, which satisfies the constraint $\widetilde{\boldsymbol{\Gamma}}_{j} \succeq  \mathcal{H}_{j}$.
Here, $\lambda_{j}$ is the maximum eigenvalue of matrix $\mathcal{H}_{j} $.
Finally, by substituting \eqref{eq:g3d}, \eqref{ap:g3h}, and $d \boldsymbol{\varphi}_j = \boldsymbol{\varphi}_j - \boldsymbol{\varphi}_j^{(t-1)}$ into \eqref{eq:hh3}, the upper-bound $\breve{g}_{\delta}(\boldsymbol{\varphi}_j|\boldsymbol{\varphi}_j^{(t-1)})$ can be given by
\begin{equation}
\begin{aligned}
\hspace{-0.3cm}\breve{g}_{\delta}(\boldsymbol{\varphi}_j|\boldsymbol{\varphi}_j^{(t-1)}) = \mathfrak{Re} \{\boldsymbol{\varphi}_j^H \boldsymbol{\Gamma}_{j} \boldsymbol{\varphi}_j \} + \mathfrak{Re} \{ \boldsymbol{\zeta}_{j}^H\boldsymbol{\varphi}_j\} + \eta_{j}, \forall j ,
\end{aligned}
\end{equation}
where
\begin{subequations}
\begin{align}
\mathbf{\Gamma}_j \triangleq& \max{\{0,\lambda_{j}\}} \cdot \mathbf{I}_N, ~ \forall j ,\\
\boldsymbol{\zeta}_{j}^H \triangleq&-2(\boldsymbol{\varphi}_j^{(t-1)})^H\boldsymbol{\Gamma}_{j} + 2\varepsilon_j^{(t-1)} (\boldsymbol{\phi}_j^{(t-1)})^H,  ~ \forall j,  \\
\eta_{j} \triangleq&g_{\delta}(\|\ln(\boldsymbol{\varphi}_{j}^{(t-1)})\|_2^2)+\mathfrak{Re} \{ (\boldsymbol{\varphi}_{j}^{(t-1)})^H \boldsymbol{\Gamma}_{j} \boldsymbol{\varphi}_{j}^{(t-1)}\}\\ \nonumber
& - 2 \mathfrak{Re} \{\varepsilon_j^{(t-1)} (\boldsymbol{\phi}_j^{(t-1)})^H \boldsymbol{\varphi}_{j}^{(t-1)}  \}
 , ~ \forall j.
\end{align}
\end{subequations}

The proof of Proposition 3 is complete.\hfill $\blacksquare$
}

%\begin{IEEEbiographynophoto}{Jane Doe}
%Biography text here without a photo.
%\end{IEEEbiographynophoto}

%\begin{IEEEbiography}[{\includegraphics[width=1in,height=1.25in,clip,keepaspectratio]{fig1.png}}]{IEEE Publications Technology Team}
%In this paragraph you can place your educational, professional background and research and other interests.\end{IEEEbiography}

\end{document}